\documentclass{emulateapj}

\newcommand{\Rstar}{\mbox{$R_\star$}}
\newcommand{\Bstar}{\mbox{$B_\star$}}

\newcommand{\mearth}{\mbox{$M_\oplus$}}
\newcommand{\msun}{\mbox{$M_\odot$}}
\newcommand{\mstar}{\mbox{$M_\star$}}
\newcommand{\mic}{\mbox{$\mu$m}}
\newcommand{\app}{\mbox{$\sim$ }}
\newcommand{\rin}{\mbox{$R_{in}	$ }}
\newcommand{\delr}{\mbox{$\Delta R$ }}
\newcommand{\lamb}{\mbox{$\lambda_0$ }}

\newcommand{\mjya}{\mbox{mJy/arcsec$^2$}}
\newcommand{\dg}{\mbox{$^\circ$}}

\begin{document}

\title {Radial Distribution of Dust Grains around HR~4796A}

\author {Z.~Wahhaj\altaffilmark{1,2}, D.~W.~Koerner\altaffilmark{2,6},
D.~E.~Backman\altaffilmark{5}, M.~W.~Werner\altaffilmark{3,6},
E.~Serabyn\altaffilmark{4}, M.~E.~Ressler\altaffilmark{3,6},
D.~C.~Lis\altaffilmark{4}}

\altaffiltext{1} 
{University of Pennsylvania, David Rittenhouse Laboratory, 
209 S. 33rd St., Philadelphia, PA 19104-6396}
\altaffiltext{2}
{Northern Arizona University, Department of Physics \& Astronomy,
Building 19, Room 209, Flagstaff, AZ 86001-6010}
\altaffiltext{3} 
{Jet Propulsion Laboratory, California Institute of Technology,
4800 Oak Grove Dr., Pasadena, CA 91109}
\altaffiltext{4}
{Division of Physics Mathematics and
Astronomy, California Institute
of Technology, Pasadena, CA 91125}
\altaffiltext{5}
{Physics and Astronomy Department, 
Franklin and Marshall College, 
P.O.  Box 3003,
Lancaster, PA 17604}
\altaffiltext{6}
{Visitors to the W. M. Keck Observatory, 
65-1120 Mamalahoa Highway 
Kamuela, Hawaii 96743 }

\begin{abstract}
We present high-dynamic-range images of circumstellar dust around 
HR~4796A that were obtained with MIRLIN at the Keck II telescope at 
$\lambda$~=~7.9, 10.3, 12.5 and 24.5~$\mu$m. We also present a new continuum 
measurement at 350~$\mu$m obtained at the Caltech Submillimeter Observatory. 
Emission is resolved in Keck images at 12.5 and 
24.5~\mic\ with PSF FWHM's of 0.37$''$ and 0.55$''$, respectively,
and confirms the presence of an outer ring centered at 70~AU.
Unresolved excess infrared emission is also detected
at the stellar position and must originate well within 
13 AU of the star. A model of dust emission fit to flux densities 
at 12.5, 20.8, and 24.5~\mic\ indicates dust grains are 
located $4^{+3}_{-2}$~AU from the star with effective size, 28$\pm$6 \mic, 
and an associated temperature of 260$\pm$40~K.

We simulate all extant data with a simple model of exozodiacal
dust and an outer exo-Kuiper ring. A two-component
outer ring is necessary to fit both Keck thermal infrared
and HST scattered-light images. Bayesian parameter estimates 
yield a total cross-sectional area of 0.055~AU$^2$ 
for grains roughly 4~AU from the star and an outer-dust disk composed
of a narrow large-grain ring embedded within a wider ring of
smaller grains. The narrow ring is
14$\pm$1~AU wide with inner radius 66$\pm$1~AU  and 
total cross-sectional area 245~AU$^2$. 
The outer ring is 80$\pm$15~AU wide 
with inner radius 45$\pm$5~AU and total 
cross-sectional area 90~AU$^2$.
Dust grains in the narrow ring are about 10 times larger 
and have lower albedos than those in the wider ring.  
These properties are consistent with a picture in which
radiation pressure dominates the dispersal of
an exo-Kuiper belt.

\end{abstract}

\keywords{circumstellar matter -- 
infrared: stars -- 
planetary systems: formation -- 
planetary systems: protoplanetary disks --
solar system: formation --
star: individual(\objectname{HR~4796A})
}

\section {Introduction}

In the last half decade, 
advances in high-resolution long-wavelength techniques have
yielded an increased number of spatially resolved images of thermal
emission from dusty disks around nearby stars 
(Greaves et al.~1998;  Holland et al.~1998; 
Jayawardhana et al.~1998 [Jay98]; Koerner et al.~1998 [K98]; 
Holland et al.~1999; Koerner, Ostroff \& Sargent~2001; 
Wilner et al.~2002; Wahhaj et al.~2003;
Weinberger et al.~2003). These  typically reveal
dust masses located at several tens of AUs
from the central star in a region analagous to the
Edgeworth-Kuiper Belt in our own solar system. In most instances,
emission is markedly reduced close to the star.
This inner clearing was previously inferred
in some cases on the basis of the spectral energy 
distribution (SED) (e.g., Jura~1998). Dust located here is most
easily detected at mid-infrared wavelengths where stellar 
photospheric radiation is relatively diminished and circumstellar
dust emission is at a maximum. If located within an 
ice-condensation radius, it can be considered
to be analogous to the solar system's zodiacal dust,
a population of grains that emanates from asteroid
collisions and the disintegration of comet tails  
(Dermott et al.~1994).
The presence of such an ``exozodiacal'' dust component
was inferred by K98 on the basis of high-resolution
($\sim0.4''$) 12 and 20~$\mu$m images of HR~4796A.

The detection of exozodiacal dust is of considerable
importance to efforts aimed at the detection of 
earth-like planets (Backman et al.~1998).
In light of a rapid dispersal timescale ($<$ 10$^4$ yr; 
Gustafson 1994), current detections imply an
origin in the collision of minor planetary bodies
and perhaps resonant trapping by larger ones. The dust
also may impact planetary detection techniques as
a source of confusing radiation (Backman et al.~1998). 
Despite their importance, detections are currently few in number.
Evidence for exozodiacal dust around HR~4796A was
recently brought into question by Telesco et al.~2000 (T00) 
and Li \& Lunine~2003 (LL03). Resolution of the 
issue is complicated by the fact that a prominent outer dust ring
provides a confusing signature at the resolution
of HR~4796A images.
Here we present additional high-resolution images of 
HR~4796A with modeling designed to better understand the
radial structure of its circumstellar dust, including
whether or not an exozodiacal dust component is
required by the data.

HR~4796A is an A0V star (Houk \& Sowell~1985) located at a 
distance of 67$\pm$3~AU ($\it{Hipparcos}$ catalog)
with an age of 8$\pm$2 million years (Stauffer et al.~1995). 
K98 and Jay98 
first resolved the disk in thermal infrared images. 
Jay98 noted that a line-cut through 
their 18~\mic\ image could be simulated 
with a ring 50 to 110~AU in extent. 
K98 simulated Keck images at higher resolution and 
demonstrated over a large range of models that 
an inner clearing was demanded by the data.
The most probable values
of disk properties yielded inner and outer
radii of 55 and 80~AU, respectively, for a 
disk inclined $\sim 73^{o}$ from face on.
K98 also found excesses at the stellar position at 
12~$\mu$m\ and 20~$\mu$m with values that implied
a dust temperature of  \app 250K and, in turn, 
a location \app 4.5~AU from the star. 
Schneider et al. 1999 (S99) imaged scattered light
from the HR~4796A outer ring  
with HST/NICMOS and confirmed the
presence of a narrow (17~AU) outer ring 
centered at 70~AU. 
T00 and 
Wyatt et al.~1999 (W99) analyzed Keck images at 10~$\mu$m and 18~$\mu$m 
and derived a radial surface density to the outer 
ring that rose exponentially from $\sim 45$~AU, doubled 
between 60 and 70~AU, and fell like $r^{-3}$ outside 70~AU with
an outer cutoff at $\sim 130$~AU. Grain sizes
were estimated to be  2-3~$\mu$m  in contrast 
to 30~$\mu$m grains found by K98. 
T00 also found that any 
excess detected in their images at the stellar position 
was consistent with their photometric uncertainties.
Augereau~et~al.~1999 presented models of the SED of
HR~4796A which 
predicted dust within 10~AU of the star. 
LL03 showed that the SED could be modeled
without a zodiacal dust component.
Sitko et al. 2000 observed that the silicate emission feature 
is very weak, suggesting that there aren't many small silicate grains 
near the star. He also estimated that only half the flux in the 
8-13.5~\mic\ range is photospheric, detecting more excess than in the 
12.5~\mic\ MIRLIN image in K98. Here we present new data and model all extant 
measurements simultaneously. These allow us to resolve
disputes about the properties of dust around HR~4796A.

\section{Observation}

\begin{figure*}[ht]
\centering
\includegraphics[width=5in]{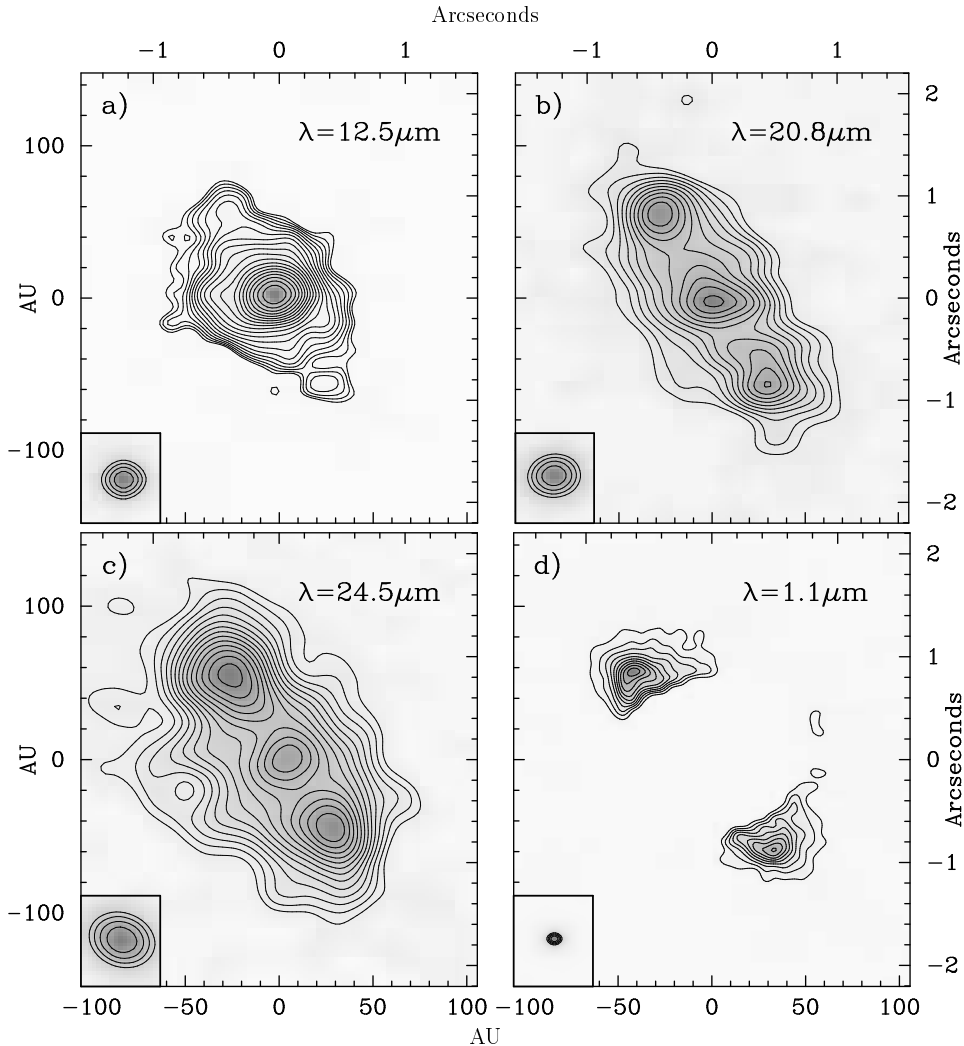}
\caption{ \small KECK/MIRLIN contour maps of HR 4796A from new observations at thermal infrared
wavelengths and the HST/NICMOS image from S99. North is aligned with the up position. 
Contour images of the PSF stars, $2$~Cen, $\alpha$~Sco and $\alpha$~Boo, 
are displayed in panels {\bf a} through {\bf c} as insets with contour intervals of 10\%, 
starting at the 60\% level. The panel {\bf d}-inset shows a TINYTIM PSF at 
the resolution (0.12'') of the HST/NICMOS observation at $\lambda$ = 1.1\mic . 
The contour levels for the {\bf d}-inset are the same as in the other panels.
{\bf (a)} Emission at $\lambda$ = 12.5 \mic\ . 
Lowest contour is at the 2$\sigma$ level (32 \mjya). Higher
contour levels are at 2$\sigma\times(10^{0.068n})$ for the $n^{th}$ contour. 
{\bf (b)} Emission at $\lambda$ = 20.8 \mic\  with 1$\sigma$
(47 \mjya) contour intervals and contours starting at the
2$\sigma$ level. {\bf (c)} Emission at $\lambda$ = 24.5 \mic\ contoured as 
in b) with 1$\sigma$ = 39 \mjya. {\bf (d)} Scattered light at $\lambda$ = 1.1 \mic\ . This
observation of HR 4796A was made with HST/NICMOS (S99) at 
1.1\mic\ . The NICMOS coronagraph was used to mask the star at the center of the image. 
Scattered light within 0.6'' of the star was deemed artifactual and is 
not shown in this image. The lowest contour is at 2$\sigma$ ( 1.8 \mjya ) 
and successive contours are drawn at 2$\sigma$ intervals.
}
\label{MIRobs}
\end{figure*}

HR 4796A  was observed on  UT 11 \& 14 June 1998, 30 January 1999, and 
3 February 1999 with the mid-infrared camera MIRLIN at the Keck II telescope.
MIRLIN's detector is a Boeing 128$\times$128 pixel, high-flux Si:As BIB 
array with a Keck II plate scale of 0.138$''$ per pixel and 17.5$''$
field of view (Ressler et al.\ 1994).  Chopping with the 
Keck II secondary mirror was carried out 
in a North-South direction at a 4 Hz rate with a throw of 7$''$,
In between chop sequences, the telescope was nodded East-West 
with a similar throw. Since thermal
IR emission from HR 4796A is confined to within a maximum length of 
5$''$, the 7$''$ chop/nod throw was sufficient for differencing
of any pair of images without overlap. 
Observations were carried out in filters centered at 
$\lambda$~=~7.9, 10.3, 12.5 and 24.5~$\mu$m with widths 0.76, 1.01, 1.16 
and 0.76 $\mu$m and for on-source integration times of 3, 6, 6, and 52 
minutes respectively. Images on  double-differenced 
frames were cross-correlated with a point spread function (PSF) to
determine a first estimate of the centroid position and were then coadded. 
The resulting image was then used in place of the PSF 
to repeat the process with improved centroiding and produce a new stacked
image. This procedure was repeated multiple times until 
there was no perceptible change in the final coadded image on successive 
iterations. The final image was sized to 64 $\times$ 64 pixels (8.8$''$ 
$\times$ 8.8$''$) to optimally window the emission from
HR 4796A and adjacent background. Infrared standards $\beta$~Leo, 
$2$~Cen, $\alpha$~Sco and $\alpha$~Boo  
were observed in the same way at 7.9 (and 10.3), 12.5, 20.8 and 24.5~\mic\ 
respectively at similar airmasses. 

To better constrain the properties of grains around HR~4796A,
a new continuum measurement was obtained on UT~May~23,~1998 
at 350~\mic\ with the SHARC linear bolometer array camera
at the Caltech Submillimeter Observatory.
Atmospheric conditions were 
good with a steady $\tau_{225}$ of 0.36. The flux calibrator was
Uranus with an angular size of 3.59$''$ and a flux density of 242~Jy on
that date. The resulting 350~\mic\ flux density of  HR~4796A was
observed to be 160$\pm$42~mJy and is listed in Table.\ 1. 

Images of HR~4796A are unresolved at $\lambda$ = 7.9 and 10.3~\mic\ with 
respective PSF FWHM of 0.48$''$ and 0.45$''$ and are not displayed here. 
Flux densities were calculated by 
comparison with images of $\beta$ Leo obtained shortly after those of 
HR 4796A. Flux densities of 307$\pm$44~mJy and 218$\pm$24~mJy at 7.9 
and 10.3~\mic\ were derived, assuming values of 10.68~Jy 
and 6.44~Jy for $\beta$~Leo at the same wavelengths (Koerner et al.\ 2000). 
Photospheric estimates 283~mJy at 7.9 \mic\ and 168~mJy at 10.3\mic , 
were derived by the fit of a model A0V star (Kurucz~1993) to measurements 
of HR~4796A at wavelengths shorter than 3~\mic . It seems that there 
is a small excess at 10.3 \mic\ and an even smaller excess or 
possibly none at 7.9 \mic .

The 12.5~\mic\ image is shown in Fig.~\ref{MIRobs}(a). The 64~$\times$~64 
image was finely gridded to 256~$\times$~256 pixels and then smoothed 
by convolution with a circular hat function of radius equal to the FWHM 
of $\alpha$~Boo (PSF star) which was observed immediately after HR~4796A. The 
resolution of the image is 0.37$''$. Peak brightnesses are calculated as 
the flux in the brightest pixel, divided by the pixel area. The central 
peak brightness was measured to be 723$\pm$44 \mjya\ while that expected 
from the photosphere is 387~\mjya. The total flux at 
12.5 \mic\ is 231$\pm$14 mJy, while 
the photospheric flux at 12.5 \mic\ is 114 mJy. The long axis enclosed by  
the 3$\sigma$ contour is 2.1$''$ (141 AU). NE and SW peaks are 
detected with total separation 1.9$''$ (127 AU). 
The NE and SW peaks are not equal in brightness, but have a 
flux ratio of 1.13.

The 20.8 \mic\ image displayed in Fig.~\ref{MIRobs}(b) is a re-reduced
version of data presented in K98. 
Shifting and adding was carried out as described above, starting
with a model of the emission derived from earlier versions of the image.
The result exhibits improved signal to noise and  
a flux density, 1.62$\pm$0.16 Jy, that is consistent with the 
previously reported 1.88$\pm$0.17~Jy. 
The final shift-and-add image was smoothed to match the 0.4$''$ 
resolution of the other images in Fig.~\ref{MIRobs}.
Emission peaks NE and SW of the 
star are separated by 1.85$''$ (124~AU) in the final smoothed image. 
The central peak is brightest and 0.93$''$ (62~AU) from the 
SW peak and the same distance from the NE peak. Peak brightness values are 
629, 668, and 623~\mjya\ at the NE, center, and SW positions, 
respectively. The photospheric flux at $\lambda$=~20.8~\mic\ is 
41~mJy, and which corresponds to a peak brightness of atmost
82~\mjya . 

The 24.5 \mic\ image is shown in Fig.~\ref{MIRobs}(c) and was reduced
in the same way as the 12.5 \mic\ image with  
$\alpha$ Boo used as the psf star. 
Image resolution was 0.55$''$.
The expected flux from the 
photosphere is 30~mJy, while the total measured flux for 
HR 4796A was 2100$\pm$170~mJy. The expected photospheric  brightness in the
peak pixel is at most 44~\mjya. Comparing this to 
the central peak brightness of 585~\mjya\, it is immediately obvious that 
most of the emission in the central pixel is non-photospheric and probably
the result of hot circumstellar dust.
The emission extends 1.68$''$ (113~AU) from the NE peak to SW peak. 
The extent of the 3$\sigma$ contour is 3.6$''$ (242~AU).
The central peak is 0.74$''$ (50~AU) from the SW 
peak and 0.94$''$ (63~AU) from the NE peak. Peak brightnesses are 637, 
617, and 585~\mjya\ at the NE, SW and central positions, respectively. 
The NE/SW brightness ratio is 637:617~=~1.04.

A very high-resolution image of light scattered by the HR 4796A
ring was obtained by S99 with the NICMOS
coronagraphic camera on HST. We have contoured
the 1.1 \mic\ image and displayed it in Fig.~\ref{MIRobs}(d) for
comparison to the thermal infrared images. The region within 0.6'' of
the star has artifactual emission from coronagraphic scattering and had been
masked off. At a resolution of 
0.12'' (8 AU),  the emission is extended 2.06$''$
(138 AU) from NE to SW peak, and exhibits a
brightness asymmetry (NE:SW) of 14.4 : 12.2 = 1.18.  The extent of 
the 2$\sigma$ contour line is much more confined than
in the thermal infrared images.

\begin{deluxetable*}{lccccc}
\tablewidth{0pc}
\tablecaption{Flux Densities for HR 4796A}
\tablehead{
\colhead{$\lambda_{eff}$} & 
\colhead{$\delta\lambda$} & 
\colhead{Flux Density} &
\colhead{Uncertainty} &
\colhead{Photosphere} &
\colhead{Reference} \\ 
\colhead{($\mu$m)} &  
\colhead{($\mu$m)} & 
\colhead{(Jy)} &
\colhead{(Jy)} &  
\colhead{(Jy)} &  
\colhead{} }
\startdata
7.9   & 0.87    &  0.307  &  0.044  & .283 &  This work                  \\
10.3  & 1.01    &  0.218  &  0.024  & .168 &  ''                         \\
12.5  & 1.16    &  0.231  &  0.014  & .114 &  ''                         \\
20.8  & 1.00    &  1.62   &  0.16   & .041 &  ''                         \\
24.5  & 0.8     &  2.1    &  0.17   & .03  &  ''                         \\
350.0 & -       &  0.160  &  0.042  & \app\ 0 &  ''                      \\
10.8  & 5.3     &  0.188  &  0.047  & .152 &  Telesco et al. 2000        \\
18.2  & 1.7     &  0.905  &  0.130  & .054 &  ''                         \\
18.2  & 1.7     &  1.100  &  0.150  & .054 &  Jayawardhana et al. 1998   \\
20.0  & -       &  1.860  &  0.186  & .045 &  Jura et al. 1993           \\
800.0 & 100.    &  0.028  &  0.0093 & \app\ 0 &  Jura et al. 1995        \\
10.1  & 5.1     &  0.270  &  0.026  & .174 &  Fajardo-A. et al. 1998     \\
10.3  & 1.3     &  0.233  &  0.024  & .167 &  ''                         \\
11.6  & 1.3     &  0.225  &  0.070  & .132 &  ''                         \\
12.5  & 1.2     &  0.253  &  0.027  & .114 &  ''                         \\
450.0 &\app\ 30 &  0.180  &  0.150  & \app\ 0 &  Holland et al. 1998     \\
850.0 &\app\ 50 &  0.0191 &  0.0034 & \app\ 0 &  ''                      \\
12.0  & 6.5     &  0.309  &  0.028  & .123 &  IRAS                       \\
25.0  & 11      &  3.280  &  0.130  & .029 &  ''                         \\
60.0  & 40      &  8.640  &  0.430  & .005 &  ''                         \\
100.0 & 37      &  4.300  &  0.340  & .002 &  ''                         \\
\enddata

\end{deluxetable*}

The images in Fig.~\ref{MIRobs} confirm the general properties of ring
emission 
inferred from previous high-resolution images (K98, S99, T00).
Flux density measurements at the stellar position also appear
to substantiate earlier evidence for a hot dust component close
to the star (K98). The latter interpretation requires careful
consideration,
however, since the highly inclined ring passes within 0.3$''$ of
the star,
and the PSF FWHM is 0.4$''$. 
It is necessary to carefully account for the
contribution of ring emission at the stellar position to identify
the
presence of hot dust near the star with certainty. In addition,
the
structure of the ring, itself, may be better ascertained with
detailed
analyses that combine information from all images. To accomplish
both
goals, we compare a variety of ring models to the Keck/MIRLIN and 
HST images.

\section {Modeling} 

Our modeling approach is designed to best interpret all available
observations of HR~4796A, including optical and thermal infrared images 
and flux densities at longer wavelengths. As a first step, we examine
individual Keck/MIRLIN images for evidence of dust close to 
the star. Then we quantitatively compare models of dust emission by
fitting them simultaneously to all available data.
To successfully deduce key disk properties in this manner,  
it is necessary to strike a balance between 
the need for a detailed physical description (including
grain shape, size and composition) and an appropriate 
number of free parameters. To best achieve this compromise, we
use a simple model of emission from a flat optically thin 
disk as described in Backman, Gillett, \& Witteborn (1992). It
will be evident from the modeling process that further complications
in disk structure, such as a flared disk, or a disk with a
thick inner edge, etc., are not required at the resolution of extant data.
According to the Backman et al. model, thermal radiation from an annulus 
of width $\it{dr}$ at distance $\it{r}$ 
from the star is

$$ f_t(r) = (1-\omega) \sigma(r) 
\varepsilon_\lambda \  B[T_p(r),\lambda] \biggl( {2 \pi r dr \over D^2} sr 
\biggr ) Jy, $$

\noindent
where  $\sigma(r)$ is the fractional surface density, $T_p(r)$ is the grain
temperature, and $D$ is the distance to HR 4796A, 67 pc. The optical 
depth is defined as $\tau(r,\nu)$ = $(1-\omega)\sigma(r)\varepsilon_\lambda$, 
where $\omega$ is the albedo.  
For moderately absorbing dielectrics and 
an effective grain radius, $a$, the radiative efficiency is 
$\varepsilon_\lambda = 1.5a/\lambda$ for $\lambda > 1.5a$ and 
$\varepsilon_\lambda = 1$ for $\lambda < 1.5a$ (Greenberg 
1979). The grain temperature for 
efficient absorbers and inefficient emitters  is $T_p(r) = 468 
(L_*/\lambda_0)^{0.2}(r/1{\rm AU})^{-0.4}$ as calculated from radiative 
balance equations (Backman \& Paresce 1993). $L_*$ is the  stellar 
luminosity in solar units. The fragmentation process caused by inelastic 
collisions between large grains results in a steady state size distribution 
described by $n(a) \sim a^{-3.5}$ (Dohnanyi 1969). It has been shown 
that the spectral energy distribution resulting from such a collisional 
cascade of grains can be approximated by the SED of a population of grains 
of a single size (Backman, Gillett, \& Witteborn 1992). The minimum grain 
radius in the distribution is $a_{min} \sim a/4$, where a is the effective 
radius of the representive grains. 

To simulate the HST/NICMOS image from S99 in Fig.~\ref{MIRobs}(d), we use the
following prescription for an emission model. The scattered-light emission from an annulus of width $\it{dr}$ at distance $\it{r}$ from the star is 

$$ f_s(r) = \omega \sigma(r) 
 \Bstar (\Rstar/ r)^2 \biggl( {2 \pi r dr \over {4 D^2} } sr 
\biggr ) Jy, $$

\noindent
where \Bstar\ and \Rstar\ are the surface intensity and radius of the star. 
It is important to note that grain size does not affect the intensity of
scattered-light emission, as grain temperature is not a factor. Therefore, the
grain radius, $a$ will not be a parameter in scattered-light modeling.

\begin{figure*}[ht]
\centerline{ \hbox{
\includegraphics[height=8cm]{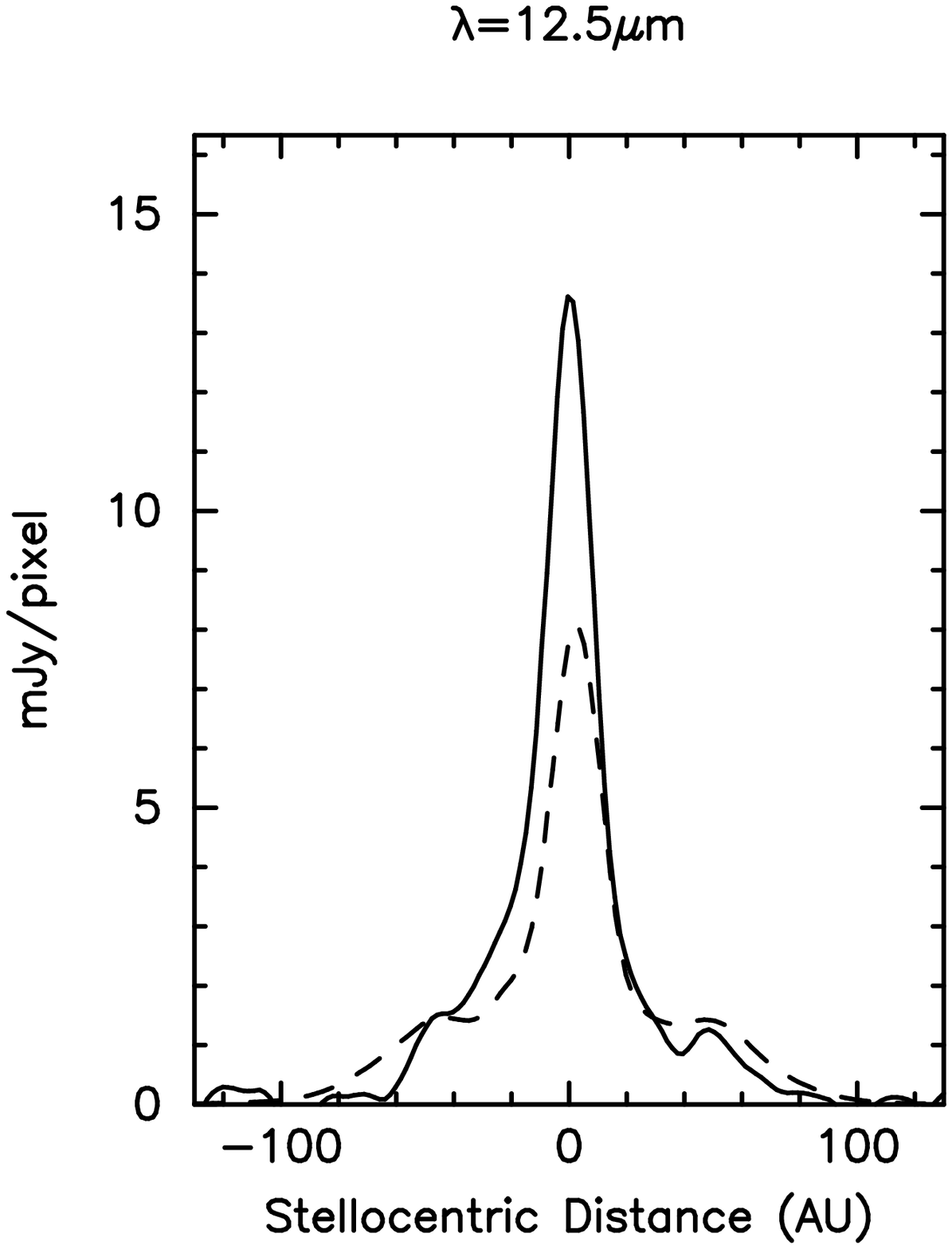}	
\includegraphics[height=8cm]{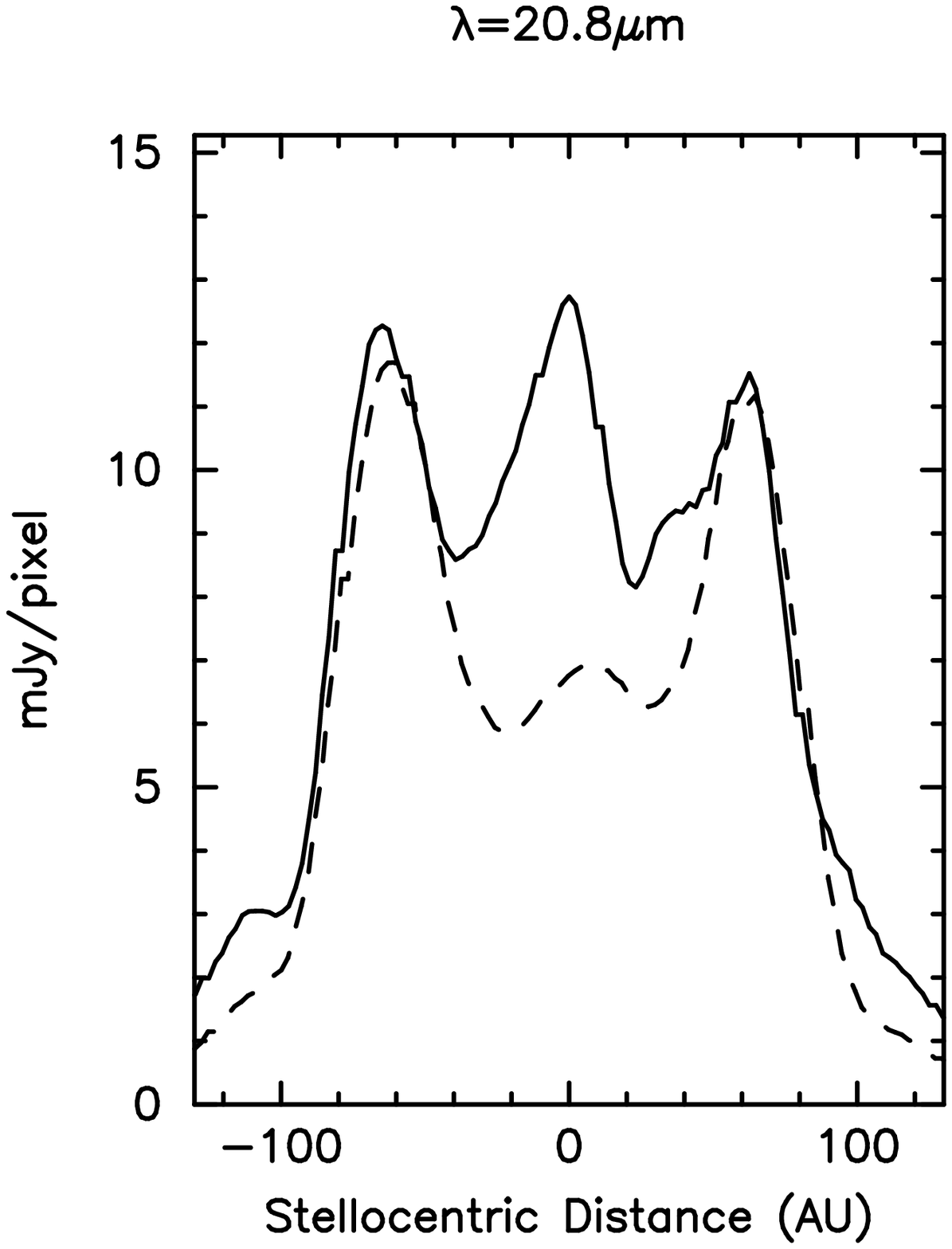}	
\includegraphics[height=8cm]{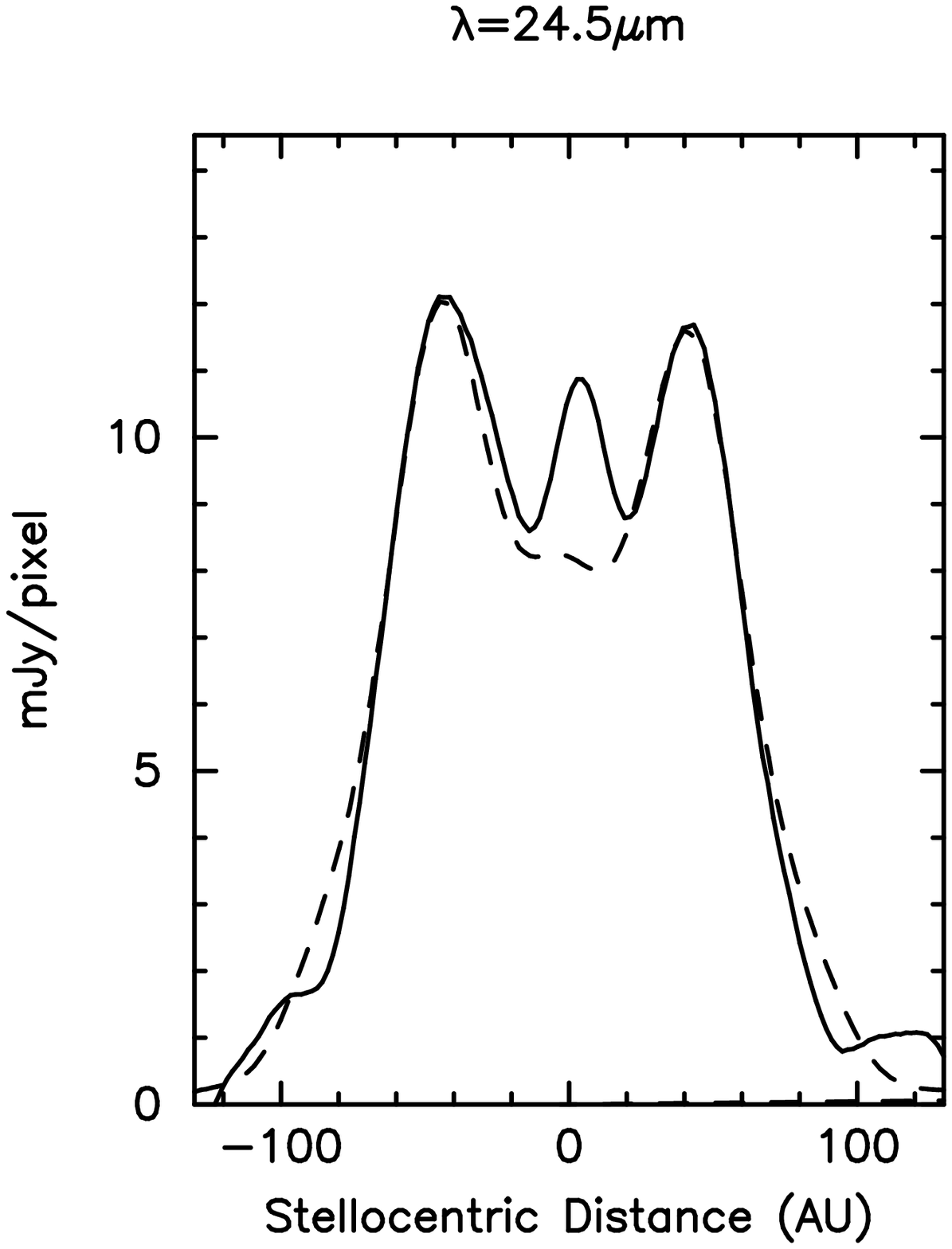}	
} }
\caption{  Models of stellar emission plus thermal emission from an outer ring 
scaled to match the outer disk's intensity are shown at 12.5, 20.8 and 24.5 \mic . 
The solid lines represent line cuts through data. The dashed lines represent line cuts 
through thermal-emission models of HR 4796A with a simple outer-ring as described in 
the text. All line cuts have a 
PA of 26\dg\ and pass through both the NE and SW peaks of emission.}  
\label{zodlc}
\end{figure*}

In keeping with simple theoretical expectations, 
models used to interpret low-resolution observations of debris 
disks often approximate the surface density, $\sigma$, 
as proportional to $r^{-\gamma}$, where $r$ is the stellocentric 
distance (Gillett 1986; Diner \& Appleby 1986; Nakano 1988; 
Artymowicz, Burrows \& Paresce 1989 and Backman, Gillett \& Witteborn 1992). 
High-resolution imaging has revealed a ring-like structure in many 
cases, however, necessitating the specification of other model 
parameters such as inner and outer radii $R_{in}$ and  $R_{out}$ and 
ring width $\Delta R$. 

To estimate photospheric emission from HR 4796A, we scale a
model of the emission from an A0V star by Kurucz (1993).
Model fitting uses all available photometric measurements of
HR 4796A at wavelengths shortward of 3 \mic.
To simulate images, the resulting photospheric
flux density is used to scale a point source
at the stellar position. As a first approximation,
we then add emission from a single-ring model and 
convolve the entire image with an appropriate PSF star. 
As demanded by the data, we then consider the addition
of multiple rings in a similar modeling process.

To calculate the range of acceptable values for
model parameters as dictated by a simultaneous
comparison to all data, we use a Bayesian approach
that assigns a relative 
probability, $e^{-1/2\Sigma_n {\chi_n^2}}$, to each model. 
Here, ${\chi_n}^2$ refers to $\chi^2$  as defined for the 
$nth$ data set. This requires a consistent way of weighting 
${\chi^2}$ from the image fits with those from SED fits.
Thus, for each image, all pixels of intensity above 3$\sigma$ are 
counted and divided by the area (in pixels) enclosed by the 
half-maximum contour of the corresponding PSF. This gives a 
measure of the number of data points represented by an image. The
${\chi^2}$s from each of the image fits is normalized using this number. 

Probabilities are calculated
for models varying over a large parameter space, and the 
relative probability of a specific parameter value 
is derived by summing probabilities for all models with 
that value. In this way, probability distributions are
built up for the values of each parameter 
(e.g. Lay, Carlstrom, \& Hills 1997). 
As a prelude to modeling the entire data set, we
first consider the question of whether or not the images
mandate the presence of material close to the star that is 
analogous to the Solar System's zodiacal dust.

\subsection{Exozodiacal Dust}

To date, consensus is lacking in published interpretations of
evidence for an inner zodiacal dust component to the HR 4796A system. 
K98 derived an infrared excess at the stellar position by
fitting a scaled PSF to high-resolution Keck images at 12.5 and 20.8 \mic\ 
and comparing the results to expected photospheric levels. 
T00 failed to detect this excess at 18.2 \mic , 
however, by carrying out a similar analysis  
on Keck images taken under non-ideal photometric conditions. 
LL03 were able to fit the Spectral Energy Distribution
(SED) without an inner dust component by varying grain composition
and using outer-disk constraints supplied by the 1.1 \mic\ HST image 
(S99).

The contradiction implied by these results reflects the difficulty
of the analysis. Thermal infrared flux calibration  
is difficult from the ground, and the inner dust component is at the 
limit of available spatial resolution. For example, T00 derived 
a total flux density 20\% lower than their previous measurement 
in the same filter at a different telescope (Jay98).
If the latter measurement is in error, it would lead to an underestimate
of the point-source excess after subtraction of flux from a photospheric
model. It is also not surprising that LL03 did not
find the case for inner dust compelling on the basis of
SED shape alone, since the function that corresponds to emission 
at a given stellocentric distance is quite broad. The dust component
inferred by K98 produces a negligible contribution to the overall
SED and can be properly substantiated only by  
high-resolution {\it imaging} data. Keck thermal 
infrared images of HR 4796A easily separate emission at the stellar 
position from that of the ring along the long axis of emission 
($\theta \sim 1''$). However, the ring is oriented 27\dg\ from edge
on, and the observations have insufficient resolution 
across the minor axis ($\theta \leq 0.25''$). 
Consequently, simple PSF fitting will be subject to 
an uncertainty from contamination by outer-ring emission. 
Here, we present a modeling approach that accounts for 
confusion from the outer ring in order to interpret 
new high-resolution images.

To make an estimate of the contribution of outer-ring 
radiation to flux detected at the stellar position, 
we first fit a simple model of outer-ring emission solely 
to the region outside of 35 AU (0.52$''$) from the star. 
Our initial model was a single 
flat ring with density $\sigma(r)$~\app~$r^{-\gamma}$,
rotational axis inclined 73\dg\ to the line of sight (S99),
and with the long axis aligned at PA 26\dg\ .  A point source
with photospheric flux density was added at the
stellar position and the entire model image convolved with a
PSF star. Ring parameters were varied in independent fits to
each of 3 images at $\lambda = 12.5,\ 20.8,$ and $24.5 \mu$m.
Fig.~\ref{zodlc} shows line cuts through both the images and best-fit
models. It is immediately evident that, {\it in each case,}
these models fail to reproduce an adequate amount of emission 
at the stellar position. This comparison strongly supports the
inference of an inner zodiacal dust component as 
concluded by K98.

\begin{figure}[ht]
\centering{
\includegraphics[width=8cm]{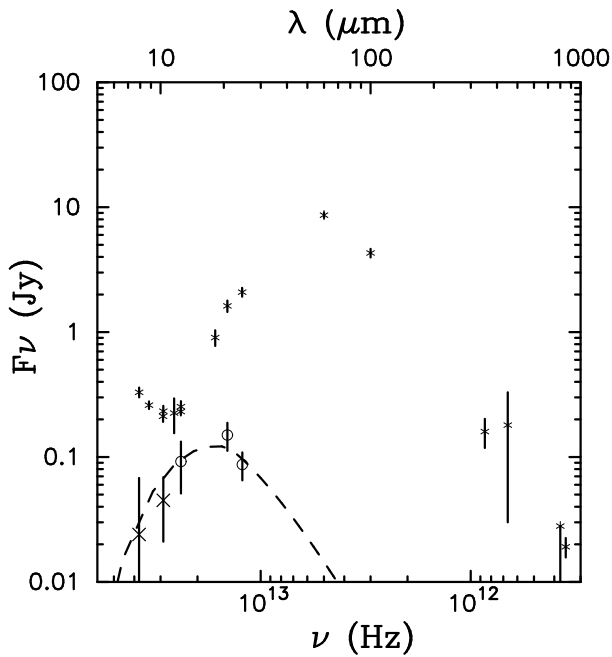}	
}
\caption{ Photometry of HR~4796A from Table~1 are denoted as asterisks. Zodiacal
excesses calculated from point-source fitting to images at $\lambda$~=~12.5, 20.8 and 24.5~\mic\ 
are shown as circles at the bottom-left corner of the figure. Estimated
flux densities from the zodiacal component at 10.3 and 7.9~\mic\ are shown as ``x''. The dashed line
represents the simulated SED of a model fitted to the zodiacal excesses. 
This model was a dust ring (T~\app~280~K) 
with radius 3.3~AU, width 
0.5~AU, grain radius 14~\mic\ and $\sigma$~=~0.0045. }
\label{zodsed}
\end{figure}

\begin{figure}[ht]
\centering{
\includegraphics[width=8cm]{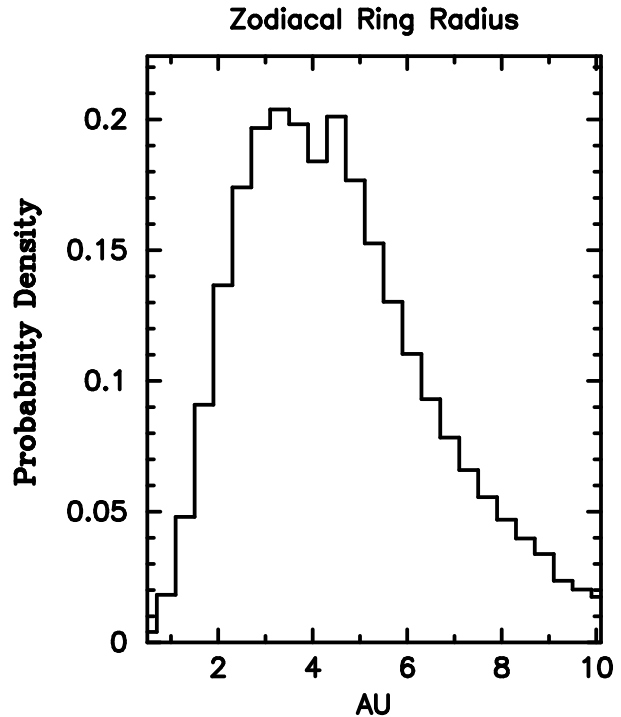}	
}
\caption{ Probability distribution for the radius of the zodiacal 
ring obtained by the Bayesian approach. The best estimate for the 
radius is $4^{+3}_{-2}$~AU.}
\label{rzodpd}
\end{figure}

We make a quantitative
estimate of the contribution of emission from inner dust by 
re-fitting our model to unmasked versions of the thermal
infrared images while varying the model flux density
at the stellar position. Expected photospheric contributions
of 114, 41, and 30 mJy at 12.5, 20.8, and 
24.5 \mic\ , respectively, were subtracted from the best-fit result 
to yield estimates of 92$\pm$41, 150$\pm$38 and 87$\pm$22 mJy for the inner
component flux densities. The results are plotted in Fig.~\ref{zodsed}. 
It is evident from Fig.~\ref{zodsed} that 
the inner component of dust contributes neglibly to the total SED
at all wavelengths. Hence, the above result is
consistent with the null result of LL03.
As a first approximation to estimate physical
properties of a zodiacal dust component, we assume grains
are located in a single ring of width 0.5 AU and vary the ring
inner radius, $r_{in},$ surface density, $\sigma$, and effective
grain radius, $a$. Best-fit values are 
$r_{in} = 4^{+3}_{-2}$~AU, $\sigma =$ 0.0046$\pm$0.001, and 
$a$ = 14$\pm$3 \mic\ with uncertainties 
determined by the Bayesian approach described earlier.
This implies an average grain temperature of 260$\pm$40~K. 
Ice grains of this size in a gas-free environment would have to be at a radial
separation of atleast 43~AU to survive sublimation for any appreciable
amount of time (Isobe 1970).
To evaluate the consistency of our 10.3 and 7.9 \mic\
photometry with this simulation, we also plot the estimated
contribution to the inner component at these wavelengths in
Fig.~\ref{zodsed}. Assuming that the outer ring SED can be approximated
by a 110K blackbody (Jura 1998) and that at 12.5 \mic\ the outer 
ring contribution is \app\ 25 mJy, we extrapolate the outer ring 
contribution to be \app\ 5 mJy at 10.3 \mic\ and and 0.2 mJy at 7.9 \mic .
Thus the inner component flux density is 45$\pm$24 mJy at 10.3 \mic\ 
and 24$\pm$44 mJy at 7.9 \mic . From Fig.~\ref{zodsed}, it is evident
that our zodiacal dust model can account for these excesses.
We emphasize the approximate nature of this estimate 
and underscore the point that the
actual dust configuration may be far more complicated.
Nevertheless, a single-ring approximation yields a radial
location as indicated in 
Fig.~\ref{rzodpd}, the probability distribution for the 
inner ring radius. The uncertainty encloses the 66\% confidence 
interval in the probability distribution.    
     
It is evident in Fig.~\ref{zodlc} that the simple power-law 
description for $\sigma(r)$ in the outer ring
fails beyond the outer-ring edge. We expect this discrepancy to
have no effect on our estimate of the zodiacal dust component.
However, we explore it below with a more detailed model to 
better understand outer-ring structure.

\subsection{An Outer Ring with Two Components}

To date, the wide variety of published imaging and photometric 
measurements of HR~4796A have not been used simultaneously to constrain 
model simulations. An oft-used single-ring model fits some of the 
individual images plausibly well. Here, however, we show that a more 
complicated model of the outer ring is required 
to fit all available data.  We first find the single-ring model 
that provides the best simultaneous fit to all data, including
Keck/MIR images, the HST-NICMOS/NIR image, and all extant flux 
density measurements of HR~4796A (Table.\ 1). 
We assume flux values for the zodiacal contribution that were
estimated in the previous section; since these contribute negligibly to
the other data, only outer-ring parameters are varied. 
For the coronagraphic HST/NICMOS image, a 
circular region of radius 0.6$''$ centered on the star is masked, so that 
only the ansae of the outer ring are considered. This accords
with S99 who point out that emission inside this 
region is largely an artifact of photospheric light scattered off
the inner edge of the coronagraphic hole. 
In image fitting, models are freely scaled and thus only fit to the 
 intensity pattern of the data. The total fluxes at the imaging  
wavelengths are fit as part of the SED.

\begin{figure*}[ht]
\centerline{
\includegraphics[height=8.5cm]{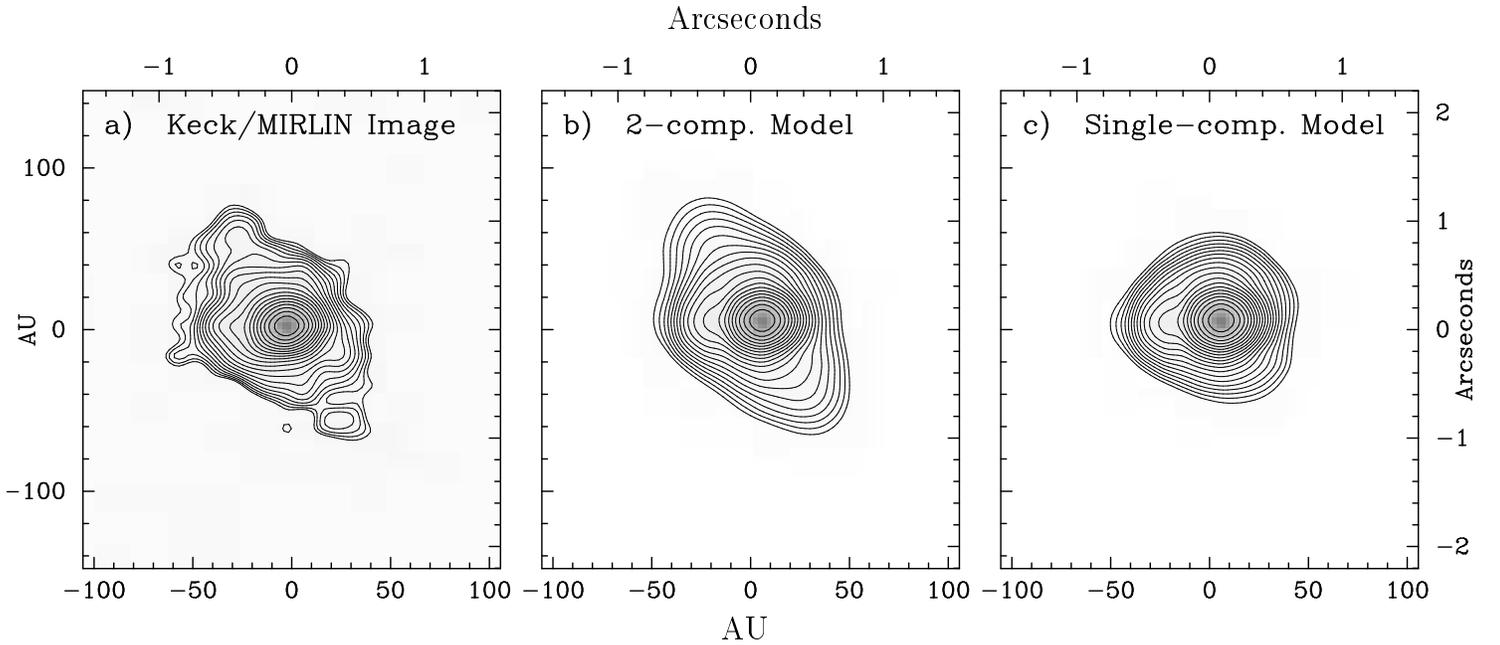}	
}
\caption{ Comparison of 12.5 \mic\ image (a) with simulations from 2-component outer-ring model (b) and a single-outer-ring model (c). Both simulations are from best-fit models to all extant data. Contour levels are same as in Fig.~\ref{MIRobs}(a)}
\label{resn5}
\end{figure*}

\begin{figure*}[ht]
\centering{ \hbox{
\includegraphics[height=10cm]{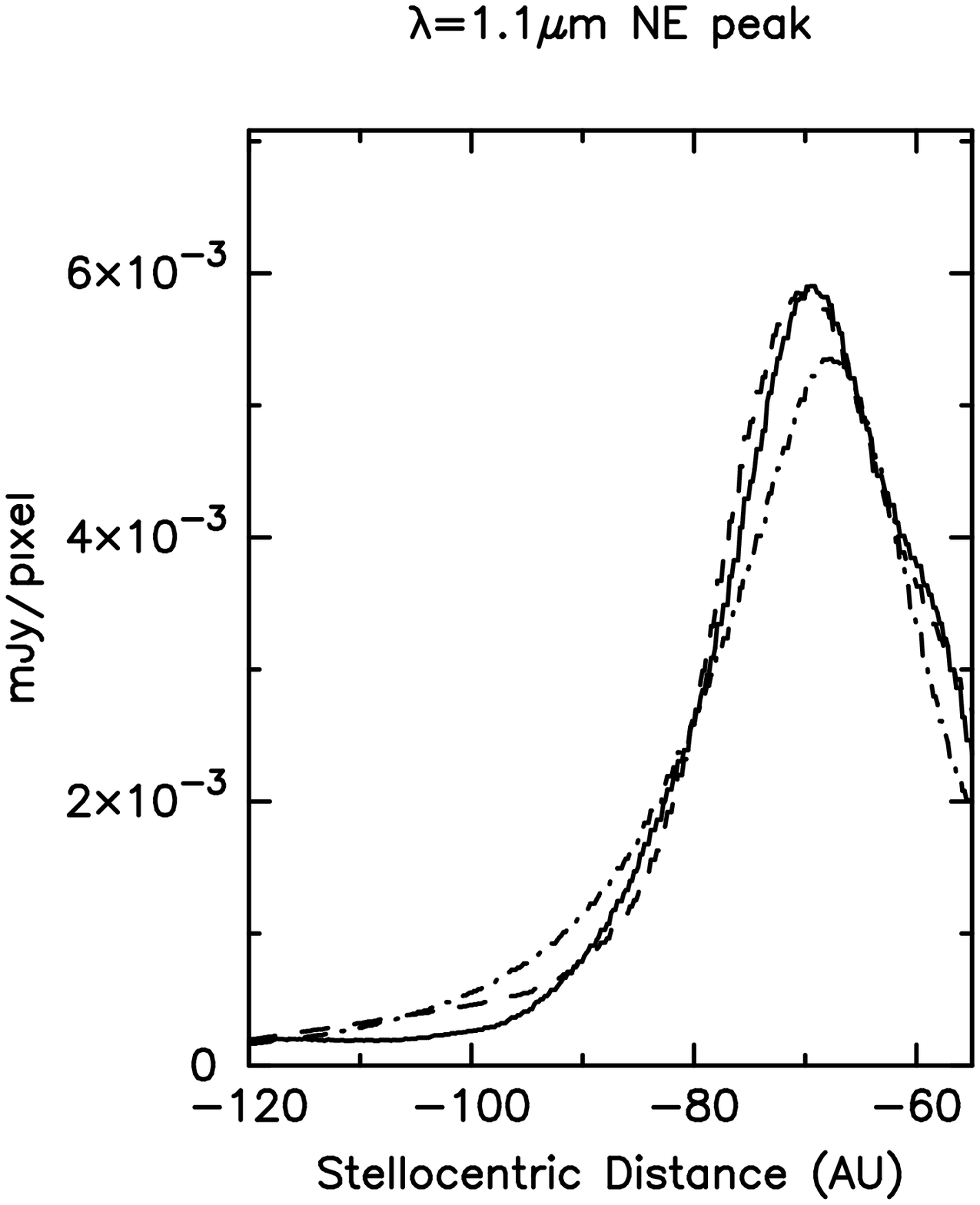}	
\includegraphics[height=10cm]{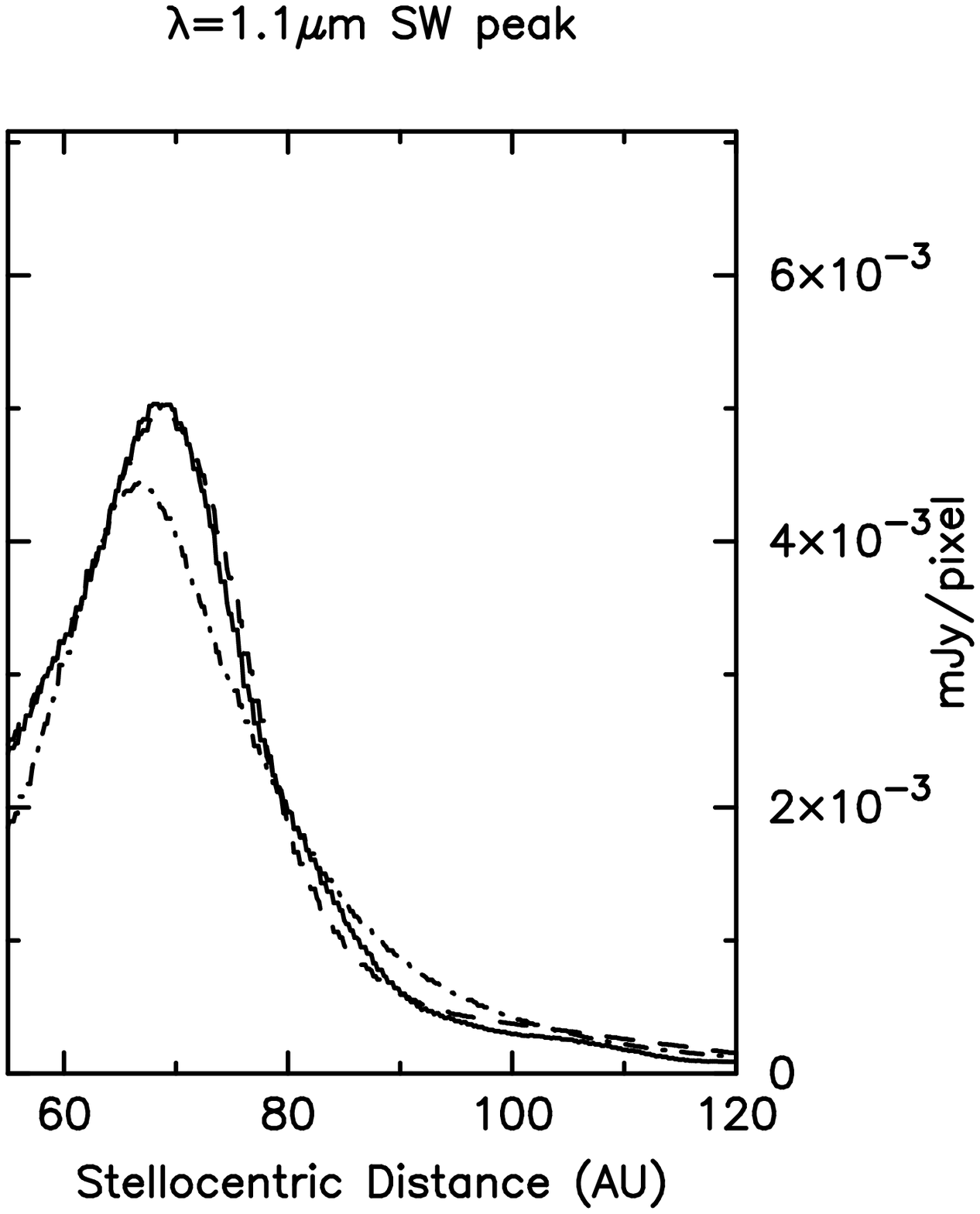}	
} }
\caption{ Strip cuts through the 1.1\mic\ HST/NICMOS image are shown as solid lines. 
Cuts through a simulated image from the 2-component outer-ring model are shown as 
dashed lines. The dash-dot-line represents the single-component model. All 8-AU-wide strip 
cuts are along a PA of 26\dg\ and pass through both NE \& SW peaks of 
emission. The NE portion is shown on the left panel and the SW portion 
is shown on the right panel. The section in between was masked by the 
NICMOS coronograph.
}
\label{resnic}
\end{figure*}
 
\begin{figure}[ht]
\centering{
\includegraphics[height=10cm]{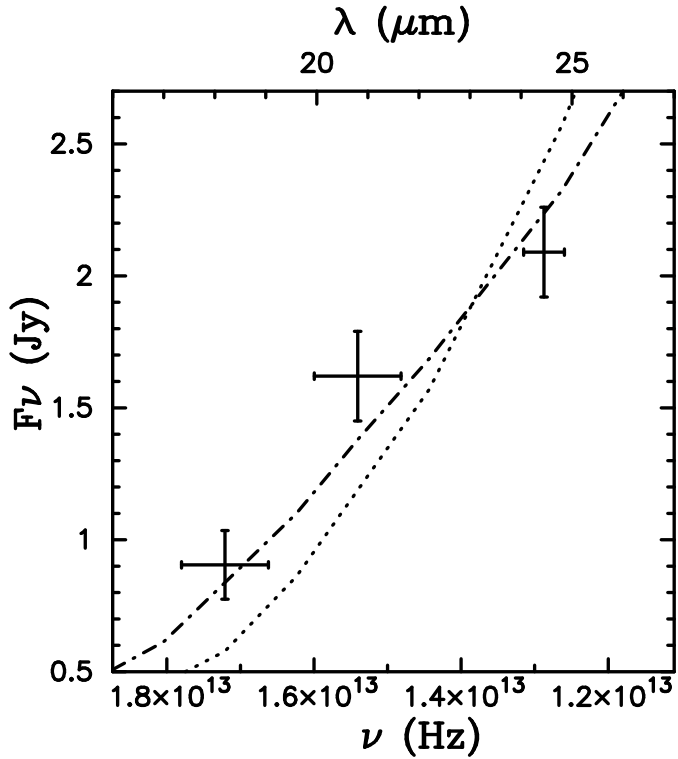}	
}
\caption{ Flux densities of HR 4796A at $\lambda$ = 18.2, 20.8, and 24.5\mic\ from Table.\ 1, shown with error bars. The dash-dot line represents the total SED from the 2-component outer-ring model. The dotted line represents the total SED from the single outer-ring model.}
\label{mirsed}
\end{figure}

\begin{figure*}[ht]
\centering{ 
\rotatebox{270}{\includegraphics[width=10cm]{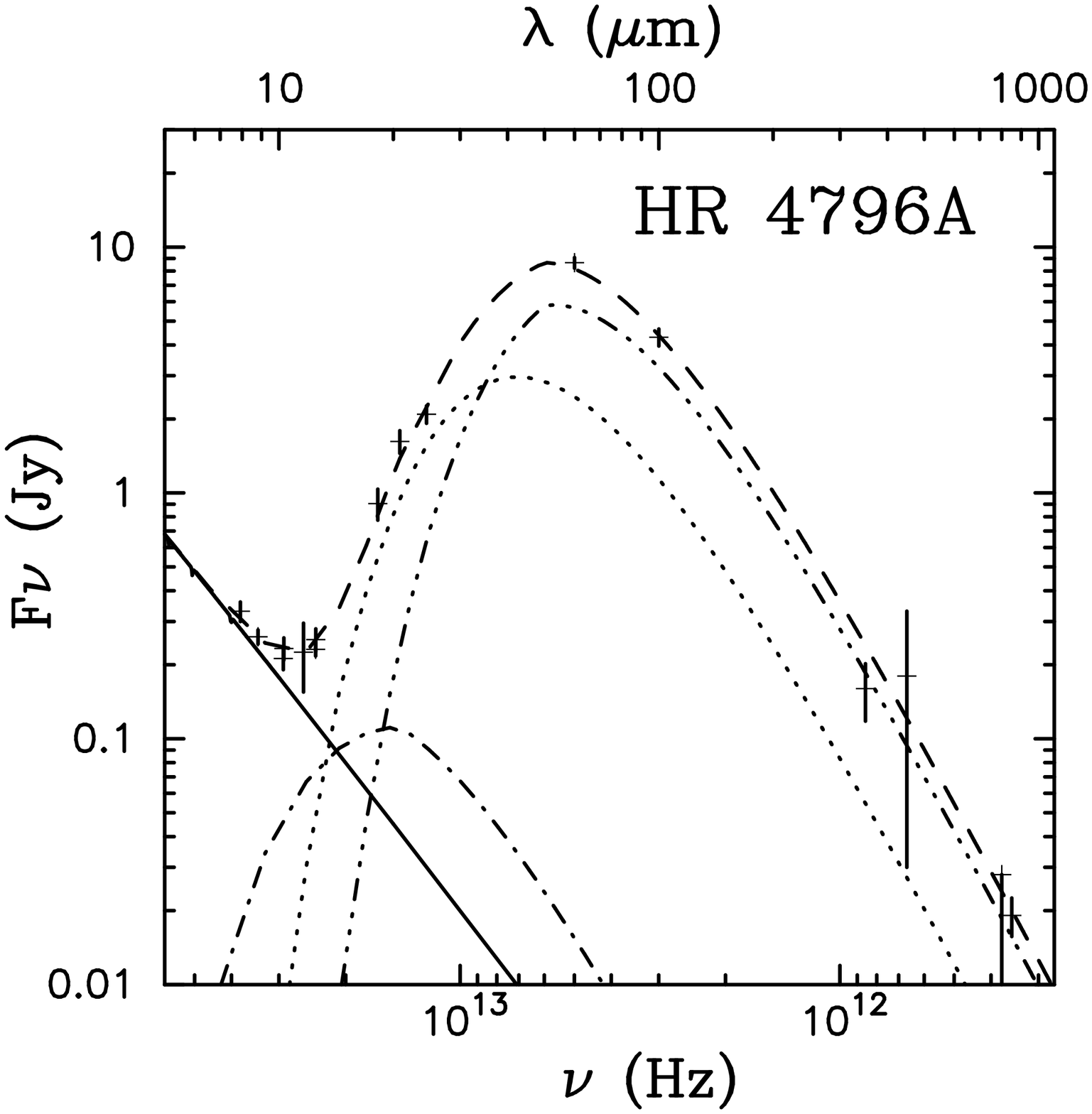}}		
}
\caption{ Flux densities of HR 4796A from Table~1. Wide-band photometry ignored where narrow-band photometry available. The dashed line represents the total SED from the best-fit 2-component outer-ring model. The solid line is the contribution from the photosphere (derived from Kurucz 1993 model). The dotted line is the contribution from the wide outer-ring component. The dash-triple-dot line is the contribution from the narrow outer-ring component. The dash-dot line is the contribution from the inner zodiacal dust component.}
\label{fnalsed}
\end{figure*}

To minimize the number of free parameters, we make one additional simplification
at the onset of detailed modeling. Searches for best-fit images revealed that
the NE ansae should have an optical depth that is 
20\% greater than the SW ansae's optical depth. So we incorporate this
asymmetry into all our models by introducing a sinusoidal azimuthal variation
into the optical depth description, such that the function has a crest at 
the NE peak and a trough at the SW peak. The amplitude of this azimuthal
variation is kept constant across all model fits, thus ridding us of one
free parameter. 

The inadequacy of the single-ring model is most evident in comparison with
images at $\lambda$~=~12.5~\mic\ and {1.1~\mic} and in a discrepancy
between model and observed fluxes at thermal infrared wavelengths.
In Fig.~\ref{resn5}(c), the best-fit single-ring model fails to simulate 
the extended emission in the 12.5~\mic\ image (Fig.~\ref{resn5}(a)). 
It is also discrepant with the MIR flux densities as evident in Fig.~\ref{mirsed}. 
Finally, emission profiles at 1.1\mic\ (solid line) in Fig.~\ref{resnic} 
are not well matched by a single-ring model (dot-dashed line) 
fit to all data. The peaks of model emission
are offset toward the star, and model flux is too high at the outer radius
and too low at the inner edge. This mismatch is consistent with the positional offset 
between peaks of scattered-light
and thermal emission in Fig.\ 6(b) of T00, an overlay of the HST/NICMOS
countour map on a 18.2\mic\ Keck/OSCIR image of HR~4796A. However, it was unclear if this
effect was simply due to the higher temperature at the inner edge of a
single outer ring. Evidence outlined above indicates 
the discrepancy in peak locations is also a consequence of 
a more complex dust distribution.
  
Some hints as to 
how the model can be improved are evident in 
fits to some of the images. A single uniform-density 
($\gamma = 0$) ring model fit to the MIR images alone gives outer-ring 
widths of \app 60~AU and grain radius \app\ 5~\mic. This is close to the grain 
radius estimates of T00, where the chief constraint is the relative 
brightnesses of the outer disks at 10 and 18~\mic s. On the other hand, 
fits to the SED alone yield a 180-AU-wide ring with roughly 16~\mic\ grains.
Smaller disk sizes, yield larger grain sizes. In fact, the Wien
side of the outer disk's SED demands small grain sizes, while the Rayleigh-Jeans
side of the outer disk's SED (the submm flux densities) demand large
grain sizes. Independant fits to the HST/NICMOS image 
yield a narrower ring, however, $\sim$ 17~AU. These discrepancies 
suggest that  the outer ring has more than one
component, and that it might be better approximated as two 
rings, each with a different characteristic grain 
size.

\begin{figure*}[ht]
\centerline{
\includegraphics[height=8.5cm]{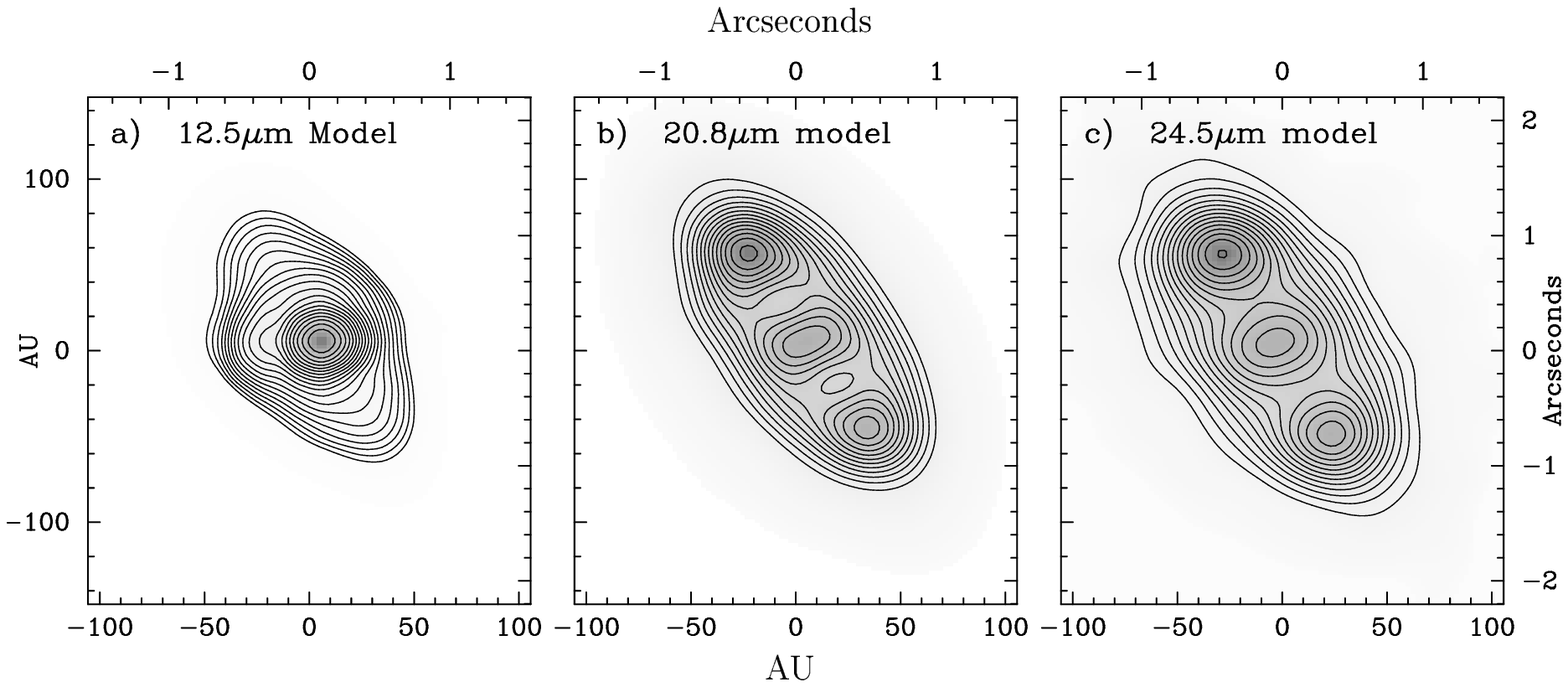}	
}
\caption{ The simulated images from the best-fit 2-component outer-ring model at the same wavelengths as in Fig.~\ref{MIRobs}.  {\bf (a)} Simulation of emission at 12.5 \mic . {\bf (b)} Simulation of emission at 20.8 \mic . {\bf (c)} Simulation of emission at 24.5 \mic . The contour levels for each figure are the same as those in Fig.~\ref{MIRobs}. 
}
\label{bestimgs}
\end{figure*}

\begin{figure*}[ht]
\centering{ \vbox{
\hbox { 
\includegraphics[height=8cm,width=8cm]{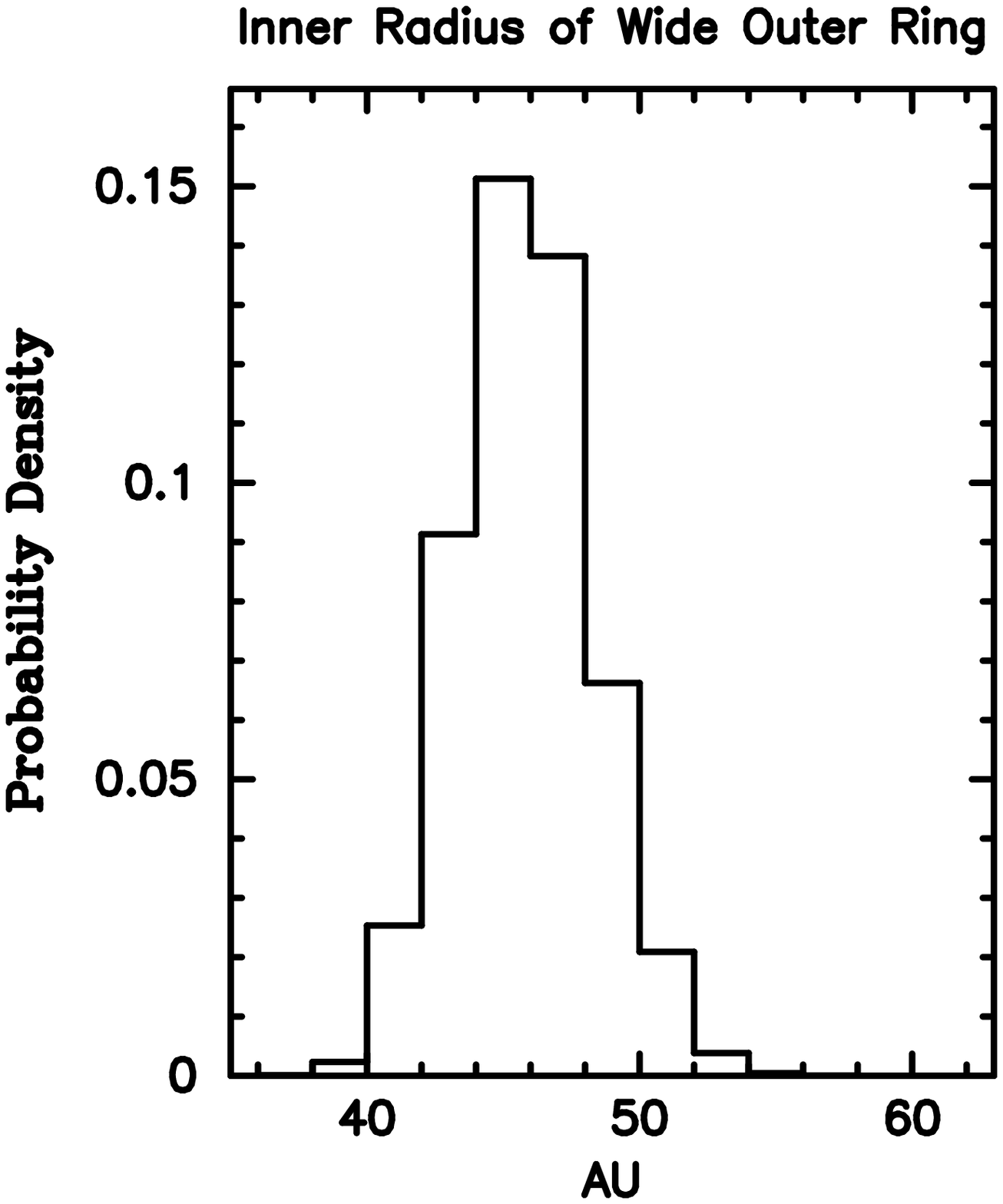}	
\includegraphics[height=8cm,width=8cm]{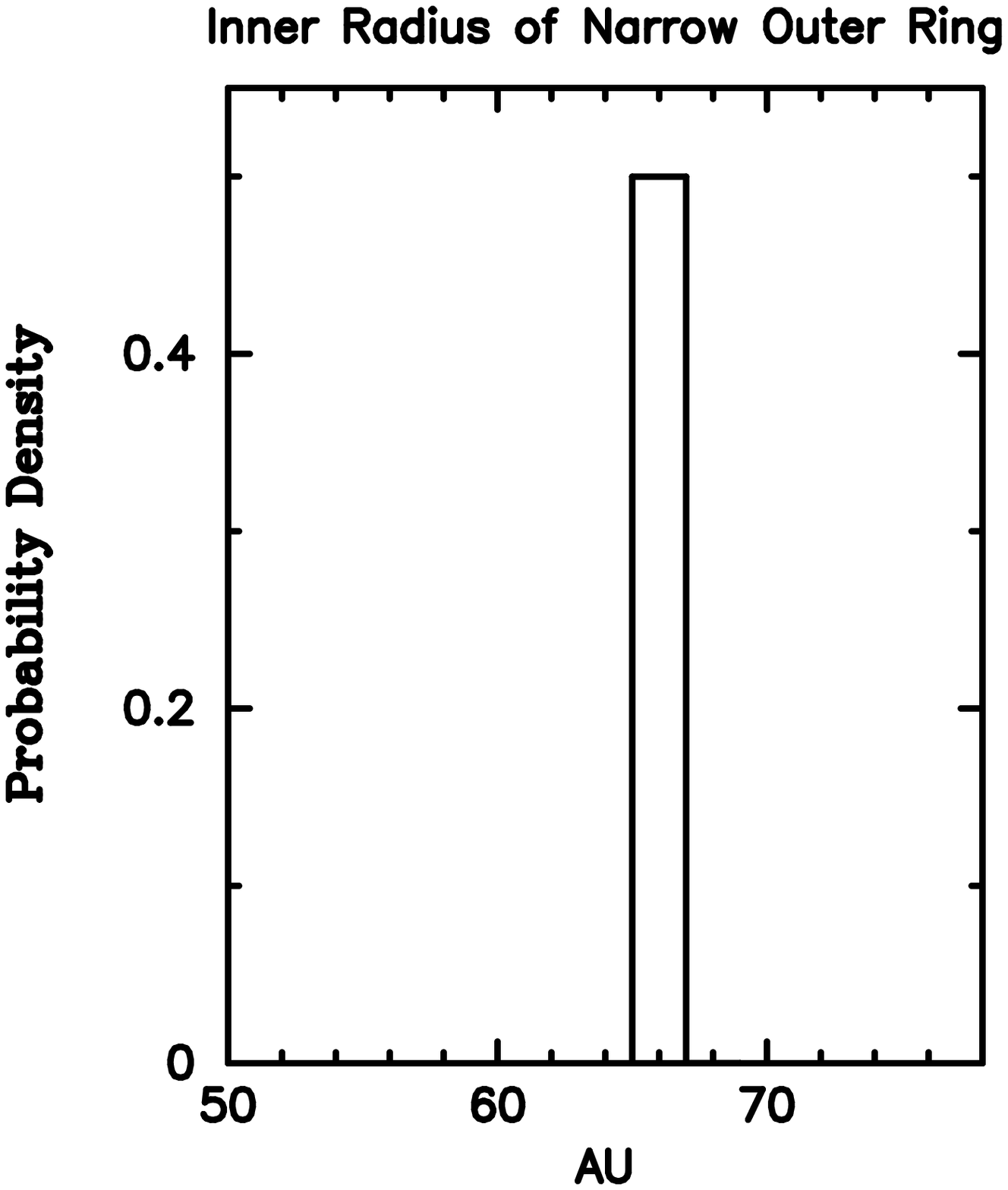}	
}
\hbox {
\includegraphics[height=8cm,width=8cm]{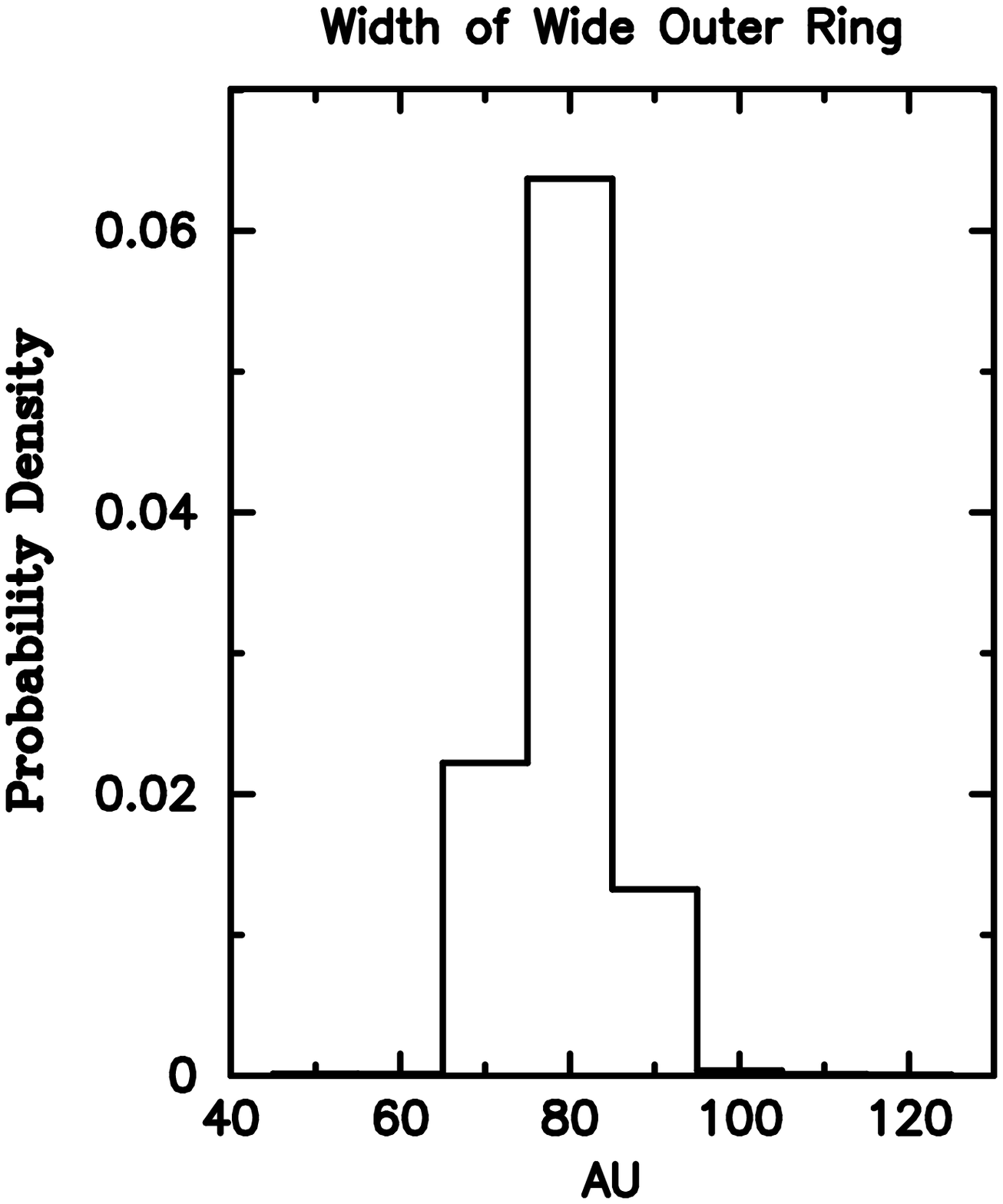}	
\includegraphics[height=8cm,width=8cm]{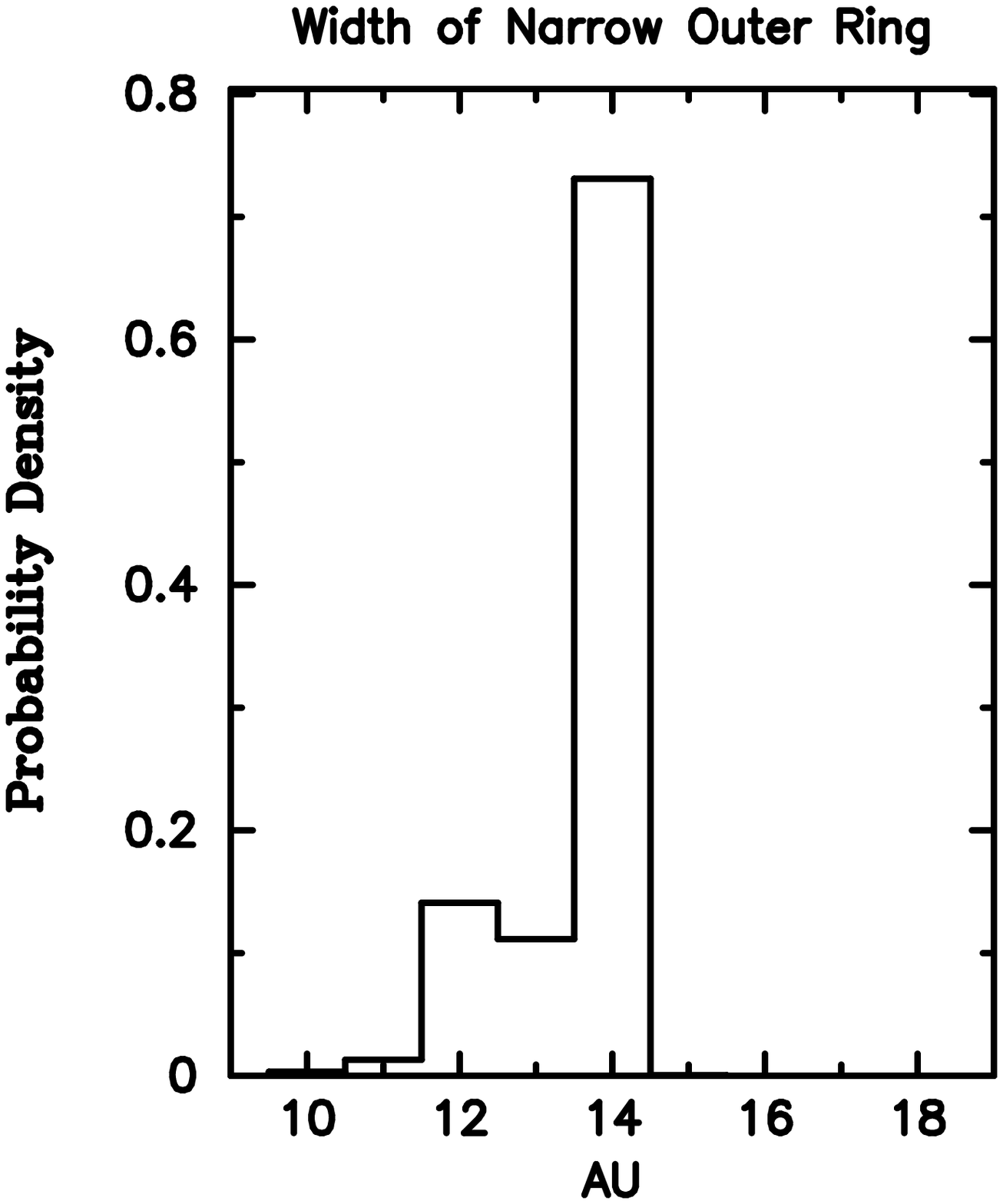}	
} }  
}
\caption{ 
Probability distributions for selected 2-component outer-ring model parameters. 
Most probable values correspond to peaks in the probability distributions. The 
uncertainties are estimated as the 66\% confidence intervals (where our distributions
are binned finely enough) or the shortest range of parameter values that 
encompasses 66\% of the total probability. The most probable values and 
uncertainties for these parameters are detailed in the text.}
\label{dists}
\end{figure*}

\begin{deluxetable}{lccccccc}
\tablewidth{0pc}
\tablecaption{The table below shows the unreduced $\chi^2$ from fits to each data set for the single-outer-ring model and the 2-component-outer-ring model. The top row of the table shows the number of data points in each data set. The next two rows show the corresponding $\chi^2$s for each model. The single-outer-ring model (1-comp) had 9 parameters, while the 2-comp had 13 parameters. According to the Bayesian Information Criterion, which takes into account the total unreduced $\chi^2$ for each model and the number of data points and free parameters, the 2-comp model is statistically preffered to a significant degree (see details in text). The $\chi^2$ values have been rounded to the nearest integer.}
\tablehead{
\colhead{$Model$} & 
\colhead{24.5\mic} &
\colhead{20.8\mic} & 
\colhead{12.5\mic} & 
\colhead{1.1\mic} &
\colhead{SED} &
\colhead{total}}
\startdata
No. data pts.   & 13 & 12 & 12 & 53 & 16 & 106 \\
2-comp $\chi^2$ & 13 & 15 & 14 & 53 & 15 & 110 \\
1-comp $\chi^2$ & 19 & 16 & 22 & 70 & 27 & 154 \\
\enddata
\end{deluxetable}

Comparisons of the 2-component outer-ring model with the 
single-outer-ring model and data are shown in 
Figs.~\ref{resn5},~\ref{resnic}~\&~\ref{mirsed}. The entire SED from the 
best-fit 2-component model is shown in Fig.\ref{fnalsed}. 
In all cases significant 
improvements are evident. However, for a more robust quantitative analysis
of which model is preferred, we can employ the Bayesian Information Criteria
($BIC$) (Liddle 2004; Mukherjee et al. 1998; Jeffreys 1961). This is
a model selection criteria which penalizes models for having extra
parameters: $BIC = -2 ln L + k ln N$, where $L$ is the likelihood function
$e^{-1/2\Sigma_n {\chi_n^2}}$, $k$ is the number of free parameters and
$N$ is the number of data points. The model with the lower $BIC$ is 
preferred. A difference of 2 between $BIC$s indicates positive evidence 
against the high-$BIC$ model, while a difference of 6 indicates strong 
evidence. The top row in Table.\ 2 shows the 
number of data points for each of the data sets 
(images and flux measuremtents).
The next two rows show the unreduced ${\chi^2}$s 
(rounded to the nearest integer) for each of the data sets
from the two best-fit models. $k$ for the 1-comp model is $9$, while
$k$ for the 2-comp model is $13$. From these numbers, the $BIC$
for the 1-component model is 196 and the $BIC$ for the 2-component
model is 171. The magnitude of this difference 
provides strong evidence that a 2-component model is 
mandated by the data.

The best-fit model (2-component outer-ring) 
is shown at MIR wavelengths in Fig.~\ref{bestimgs}. 
Single-ring best-fit parameters are: 
\rin$=64\pm4 AU$, \delr$=80\pm6 AU$, $\gamma = 4.88\pm0.2$, \lamb$=31\pm3$ \mic\ 
($a = 21$ \mic ), $\sigma = 3.3\pm0.2 \times 10^{-3}$ and albedo$=0.3\pm0.02$. 
Uncertainties were estimated through the Bayesian approach. The 2-component outer-ring model parameters are: 
$R_{in1}=45\pm5 AU$, $\Delta R_1=80\pm15 AU$, 
$\lambda_{01}=5\pm2$ \mic\ ($a_2 = 3.3$ \mic ), $\sigma_1 = 2.1\pm0.2 \times 10^{-3}$ 
(implying a total cross-sectional area of 90 AU$^2$)   
and albedo, $\omega_1 = 0.35\pm0.03$,
$R_{in2}=66\pm1 AU$, 
$\Delta R_2=14\pm1 AU$, $\lambda_{02}=38\pm5$ \mic\ ($a = 25$ \mic ), 
$\sigma_2 = 3.8\pm0.25 \times 10^{-2}$ 
(implying a total cross-sectional area of 245 AU$^2$)
and albedo, $\omega_2 = 0.18\pm0.02$. 
The $\gamma$ parameter was not used for either ring. The probability 
distributions, obtained by the Bayesian approach, for 
some of these parameters are shown in Fig.~\ref{dists}.

\section {Discussion} 

Analysis of multi-wavelength HR 4796A images and
flux densities yields 
a refined estimate of the properties of its circumstellar 
dust as follows. Radial structure of the outer ring is best 
approximated by two components: a narrow ring of $\sim$50 \mic\ 
grains between 66 and 80 AU and a surrounding wider ring  
of $\sim$7 \mic\ grains stretching from 45 AU to about 125 AU. 
The presence of excess emission at the stellar position is confirmed. 
It is unresolved by a 0.37'' FWHM PSF (corresponding to a 12.5 AU 
radius at the 67 pc distance of HR 4796A) and has a temperature
of \app~260 K. Bayesian parameter estimation suggests this dust is
\app\ 4 AU from the star with a total grain 
cross-sectional area of 0.055 AU$^2$.

\subsection {Exozodiacal Dust}

Evidence for warm exozodiacal dust was first 
identified for HD~98800 (Zuckerman \& 
Becklin 1993). This K-dwarf quadruple system of two spectroscopic 
binaries (Torres et al.\ 1995) exhibits one of the strongest dust
signatures of any in the IRAS catalog. Its infrared
excess peaks at 25 $\mu$m and is associated entirely
with the northern spectroscopic binary, HD 98800 B 
(Koerner et al.\ 2000). Modeling of the spectral energy
distribution of the dust suggests a ring of particles
in the few-micron size range with an inner radius of 
a few AU and an outer radius that may stretch will
into a giant-planet zone. Subsequent detections of
warm excess from the main-sequence A3 star, $\zeta$ Leporis,
have also been intepreted as orginating from collision in 
an asteroid belt (Chen \& Jura 2001). 

The origin of a warm component to circumstellar dust 
around HR 4796A may be similar to that of our own Solar System.
Zodiacal  dust is supplied predominantly by the asteroid belt, 
Jupiter Family comets (JFCs), long-period comets,
and perhaps by dust from the Kuiper-belt and the Oort Cloud 
(Flynn 1994, Dermott et al. 1992, 
Durda \& Dermott 1992, Liou \& Dermott 1993). 
At 1 AU, about 45\% of the dust in the ecliptic plane 
is believed  to come from the asteroid belt and JFCs. 
Within a sphere of radius 1 AU, however, 
89\% of the dust is from comets. The fractional surface density
 of zodiacal dust grains is $\sigma(r)$ \app\ $r^{-1.45}$
(Hahn et al. 2002) with grain sizes as large as 100 \mic\      
(Grogan et al. 2001).

An asteroid belt may be the source of  
exozodiacal dust grains around HR 4796A. Their 
total grain cross-section of $\sim 0.055\ AU^2$ and
effective grain size of 28 \mic\ yields a minimum mass of 
$\sim 10^{-3}$ \mearth, under the assumption that the grain density is 
given by $\rho$=2500$kg/{m^3}$ (Jura 1998). Ice sublimation occurs 
instantaneously at 4~AU from the star (Isobe 1970), so these grains are
almost certainly refractory. The blow-out size due to
radiation pressure is 7 \mic\ (calculated according to 
Backman \& Paresce 1993, Gustafson 1994) and indicates this 
is not an effective dispersal mechanism for the grains
we detect. Due to relatively high fractional surface density 
($\sigma$ = 0.0046), the collisional lifetime is shorter than 
the Poynting-Robertson lifetime even at 4~AU. 
Grains are primarily being destroyed by collisions,
but probably also originate in a cascade of collisions from 
larger bodies. Assuming a grain size distribution, 
$n(a)$ \app\ $a^{-3.5}$, and 1000 km planetesimals for the  
largest bodies, the total mass of orbiting material 
is \app 0.6 \mearth .

Another possible source for HR 4796A's exozodiacal dust is 
a more distant asteroid or exo-Kuiper belt. Grains 
may spiral in toward the star under the influence of 
Poynting-Robertson drag but be trapped by resonant interactions 
with terrestrial planets. The dust may come only from more tenuous
parts of the disk where collisional destruction is not the dominant 
dispersal phenomenon. For our Solar System, 
however, Liou et al. (1996) showed that Kuiper belt grains in the range 
9 to 50~\mic\ are unlikely to survive collisions with interstellar 
grains (as opposed to interplanetary grains) on their way to the Sun, 
and 80\% of all migrating
grains are likely to be thrown out of the Solar System
by interactions with the giant planets. However dust can be trapped
in planets interior to an asteroid belt, and in these cases grains
roughly 30~\mic\ or greater in size seem to be the best candidates
(Jackson \& Zook 1992). Large asteroid grains can even be injected
into interior resonances with exterior planets due to the initial outward thrust
of radiation pressure upon release from a parent body. As they leave these
resonances they usually have lower eccentricities and orbital inclinations
that render them susceptible to capture in resonant orbits 
with interior planets.

Comets that travel close to the star may
provide a source of exozodiacal dust grains around HR~4796A.
The cometary component
of zodiacal dust is thought to be highly variable according 
to simulations done by Napier 2000
and may increase the dust creation 
rate from a 100 to 1000 fold within 
periods of 100,000 to 10,000 yrs. 
Particles that are  most likely to be trapped in our own Solar System
have sizes between 10 and 100 \mic\ (Jackson \& Zook 1992). Dust with
low eccentricities and low orbital inclinations are again more likely
to be captured in resonant orbits. It should be noted that
while the radiation pressure blow-out size for the Solar System is
roughly 0.9 \mic , it is 7 \mic\ for HR~4796A. The 
optical depth of the zodiacal dust at 25 \mic\ 
is roughly $10^{-7}$ when measured along the ecliptic 
(Spiesman et al.\ 1995). For HR 4796A, assuming similar
dust dispersal times scales, this would imply a dust 
production rate roughly $10^{4}$ times higher 
than expected for the Solar Sytem. 
Grain velocities also differ in the two systems. Thus 
it is best to undertake system-specific
simulations before speculating too much about the 
origin of exozodiacal dust grains.
Regardless of origin, the main explanation for the 
persistence of dust in the exozodiacal
region is probably resonance trapping by interior planets 
as found in the Solar System.

\subsection {Properties of HR 4796A's Outer Ring}

A cold, massive outer ring of dust particles is more readily
detected around stars other than the Sun. Examples other than
HR 4796A include Fomalhaut (Holland et al.\ 1998); 
$\epsilon$ Eri (Greaves et al.\ 1998), and 
Vega (Holland et al.\ 1998; Koerner et al.\ 2001; Wilner et al.\ 2002).
However, such a structure is more difficult to detect in our
own Solar System where confusion from the inner zodiacal
dust interferes with emission from outer dust 
grains (Backman et al.\ 1998).
The discovery of a sizeable aggregation of Kuiper Belt Objects
(see review by Luu \& Jewitt 2002) implies an associated
population of dust grains (Backman et al.\ 1995). Larger grains
are likely to be better retained in the Kuiper Belt region by  
mean motion resonances while smaller grains
diffuse  under the influences of P-R drag, solar-wind drag, 
and radiation pressure (Holmes et al.\ 2003).

The HST/NICMOS image of HR 4796A at 1.1 \mic\ 
establishes that most of the scattered light around 
HR 4796A arises within a 17 AU ring centered at 70 AU
(S99).  A wider low-density component is 
also needed to fit Keck/MIRLIN images of thermal emission.
The spectral energy distribution can be fit
by a single narrow ring, but only if 
the large-grain contribution to the
size distribution is boosted significantly.  
LL03 used $n(a) \sim a^{-2.9}$ to do this, 
but their models fail to match the surface brightness 
distribution in thermal infrared images. 
Inclusion of two distinct effective grain sizes 
resolves an apparent inconsistency in the interpretation
of different data sets. Thermal infrared
images modeled by T00 required
small grains (\app\ 3 \mic ), but a fit by K98 that included
an upper limit to the flux density at 800 \mic\ required
the presence of large grains (\app\ 30 \mic ) to fit the 
Rayleigh-Jeans side of the SED. The latter 
result is confirmed here by fits which include
more recent detections of emission at 450 and 800 \mic\ 
(Holland et al. 1998) and a new measurement at 
350 \mic . A two-part ring with separate grain-size 
distributions is necessary to reproduce observations at
{\it all} wavelengths.

Properties derived here for HR 4796A's outer ring are
consistent with theoretical predictions for both our Solar System's 
outer dust and for exo-Kuiper grains in general.
The collision of large planetesimals is expected to 
produce a cascade of smaller-sized particles with number distribution
$dn(a)/da \sim a^{-3.5}$ (Dohnanyi 1969).
Subsequent grain evolution is affected largely
by radiation pressure; Poynting-Robertson drag is a relatively
small effect. Radiation pressure effects can be parametrized by $\beta$:
$$ \beta(D) = (1150/ {\rho D}) (L_\star / L_\odot ) (M_\odot / M_\star ) $$
\noindent
where $D$ is grain diameter in $\mu$ms, and $\rho$ is grain density in 
SI units (Gustafson 1994, W99).
Appropriate values for HR 4796A are \mstar = 2.5 \msun (Jura 1998) 
and $\rho$=2500 $kg/{m^3}$ and yield $\beta(D) = 3.5/D$. 
Large particles with $D > 35$ correspond to $\beta < 0.1$ and
will remain in original orbits close to parent bodies.  
Grains in the size range 35 \mic\ $> D >$ 7 \mic\ are 
$\beta$ critical with values $ 0.1 < \beta < 0.5$.
These experience orbital evolution which extends their 
distribution both inwards and outwards. 
The smallest grains ($D < 7$ \mic\ and $\beta > 0.5$) have been
called  ``$\beta$ meteoroids'' and are blown out of the system
on hyperbolic orbits. 

This picture of grain evolution provides a natural explanation
for the properties of the outer ring around HR 4796A.
In this scenario, a large-grain ring centers on a system of
planetesimals surrounded by an outer region of smaller
grains diffused under the influence of radiation pressure.  
Large grains in the narrow ring have 
$\beta <  0.1$, $D > 35$ \mic\ or $D$ \app 50 \mic\ in our model, so they 
stay close to parent bodies.  Smaller grains  
are ``$\beta$ critical'' or ``$\beta$ meteoroids,'' since their size is 
so close to 7 \mic\ ($D$ \app 6.6 \mic ) threshold. However,  ``$\beta$ 
critical'' is more likely, since ``$\beta$ meteoroids'' have a ``blow-out'' 
lifetime of order 100 yrs, while ``$\beta$ critical'' dust has a 
collisional lifetime of order 1000 yrs.  This idea is supported
by comparison with expected size distributions. The large- and small-grain
populations have a diameter ratio of about 7:1 and cross-sectional 
areas $90\ AU^2$ for small grains and $245\ AU^2$ for 
large grains. So the number of grains is approximately in the ratio 20:1 
for small:large grains. In contrast, the initial ratio for
a collisionally generated size distribution should be 
$(7/50)^{-3.5} \sim 1000:1$. This underabundance of small grains 
agrees well with the hypothesis that 
many ``$\beta$ meteoroids'' have already left the system.

\subsection {Furthur Work}

The modeling approach presented in this work 
was designed to determine the simplest disk morphology 
required by observations with a minimum number of 
underlying assumptions. The resulting dust distribution
suggests significant physical interpretations.
These should be tested with a more complex analysis
that uses a more realistic representation of dust
grain properties, including physically meaningful
size distributions and plausible assumptions about 
grain composition. The latter should, in turn, be 
coupled to models that simulate the dynamical evolution
of dust grains under the influence of planetary bodies
in the HR~4796A system.  

Further progress in observing HR 4796A's radial dust distribution will be 
difficult until a next generation of large-aperture telescopes comes on line,
since Keck/MIRLIN images are already close to the diffraction limit of a 10-m
telescope. 
ALMA may provide sub-millimeter images with resolution improved by a factor of 
5. The next generation telescopes like the 30 meter 
``Giant Segmented Mirror Telescope'', 
the ``Thirty Meter Telescope'' 
and the 100 meter ``Overwhelmingly Large Telescope'' 
will have also improved resolution and
 greatly enhanced sensitivity, and high-dynamic range techniques that are
designed to search for planets may provide more detail on reflected light. 
Meanwhile, progress in observations of exozodiacal and exo-Kuiper dust configurations, may be
better advanced by studying systems that are closer than the 67 pc distance to
HR 4796A. The enhanced sensitivity of {\it Spitzer Space Telescope} will 
identify nearby debris-disk systems for which spatial resolution will be 
improved. Nulling interferometers (Keck and LBT) will also bring enhanced capability
to study their inner dust regions.  Ultimately, coordinated planet searches 
will be able to verify the relationship between circumstellar dust signatures and the
presence of a planetary system.

Portions of this work were carried out at the Jet Propulsion Laboratory,
operated by the California Institute of Technology under a contract 
with NASA. Data presented herein were obtained at
the W.M. Keck Observatory, which is operated as a scientific 
partnership among the California Institute of Technology, the 
University of California and the National Aeronautics and Space 
Administration. The Observatory was made possible by the generous 
financial support of the W.M. Keck Foundation. The authors wish also 
to recognize and acknowledge the very significant cultural 
role and reverence that the summit of Mauna Kea has always had within the 
indigenous Hawaiian community.  We are most fortunate to have the 
opportunity to conduct observations from this mountain.

\vfill
\eject


\begin{references}

\reference{arty88} Artymowicz, P., Burrows, C., Paresce, F.\ 1989, \apj, 337, 494
\reference{aug99} Augereau, J.C., Lagrange, A.M., Mouillet, D., Ménard, F.\ 1999, A\&A, 350, 51
\reference{bgw92} Backman, D.E., Gillett, F.C., \& Witteborn, F.C., 1992,\apj, 385, 670
\reference{bp93} Backman, D.E., \& Paresce, F., 1993, in Protostars \& Planets III, (ed. E.H. Levy \& J.I. Lunine), Tucson: University of Arizona Press, p. 1253
\reference{bak95} Backman, D.E., Dasgupta, A. \& Stencel, R.E.\ 1995, \apj, 450, 35
\reference{bak97} Backman, D.E., Caroff, L.D., Sandford, S.A., Wooden, D.H.\ 1998, Exozodiacal Dust Workshop;
 Conference Proceedings, (NASA Ames Research Center)
\reference{cj01} Chen, C. H. \& Jura, M. \ 2001, \apj, 560, 171
\reference{faj98} Fajardo-Acosta, S.B., Telesco, C.M., Knacke, R.F.\ 1998, \aj, 115, 2101
\reference{der94} Dermott, S.F., Jayaraman, S., Xu, Y.L., Gustafson, B.A.S., Liou, J.C \ 1994, Nature, 369, 719
\reference{der92} Dermott, Stanley F., Durda, Daniel D., Gustafson, Bo A. S., Jayaraman, S., Xu, Y. L., Gomes, R. S., Nicholson, P. D., In Lunar and Planetary Inst., Asteroids, Comets, Meteors 1991, 153
\reference{da86} Diner, D.J. \& Appleby, J.F.\ 1986, Nature, 322, 436
\reference{dd92}  Durda, D.D., Dermott, S.F., American Astronomical Society, 24th DPS Meeting, 11.04, Bulletin of the American Astronomical Society, 24, 951
\reference{d69} Dohnanyi, J., 1969, \jgr, 74, 2531
\reference{flynn94}  Flynn, Abstracts of the 25th Lunar and Planetary Science Conference, 1994., p.379
\reference{g98} Greaves, J.S., Holland, W.S., Moriarty-Schieven, G., Jenness, T., 
Dent, W.R.F., Zuckerman, B., McCarthy, C., Webb, R.A., Butner, H.M., Gear, W.K. \& 
Walker, H.J.\ 1998, ApJ, 506, 133
\reference{gre79} Greenberg, J.M.,\ 1979, Infrared Astronomy, Proceedings
of NATO Advanced Study Institute, ed. G. Setti \& G.G.Fazio 
(Dordrecht:Reidel), 51
\reference{gil86} Gillett, F.C.\ 1986,  Light on dark matter, Proceedings of the First 
Infra-Red Astronomical Satellite Conference, Noordwijk, Netherlands, June 10-14, 1985 
(A87-11851 02-90). Dordrecht, D. Reidel Publishing Co., 61
\reference{grog01} Grogan, K., Dermott, S.F., Durda, D.D.\ 2001, Icarus, 152, 251
\reference{gus94} Gustafson, B.A.S.\ 1994, Annual Review Of Earth And Planetary Sciences, 22, 553
\reference{hahn02} Hahn, J.M., Zook, H.A., Cooper, B., Sunkara, B.\ 2002, Icarus, 158, 360
\reference{hol98} Holland, W.S., Greaves, J.S., Zuckerman, B., Webb, R.A.,McCarthy, C., Couldon, I.M., Walther, D.M., Dent, W.R.F.,Gear, W.K., \& Robson, I., 1998, \nat, 382, 788
\reference{hol99} Holland, W. S., Robson, E. I., Gear, W. K., Cunningham, C. R., Lightfoot, J. F., Jenness, T., Ivison, R. J., Stevens, J. A., Ade, P. A. R., Griffin, M. J., Duncan, W. D., Murphy, J. A., Naylor, D. A.1999 MNRAS, 303, 659
\reference{holmes03} Holmes et al.\ 2003, \apj, 597, 1211
\reference{hs85} Houk, N.; Sowell, J. \ 1985, BAAS, 17, 878
\reference{iso70} Isobe, S.\ 1970, Publications of the Astronomical Society of Japan, 22, 429
\reference{jh89} Jackson, A.A., Zook, H.A.\ 1992, Icarus, 97, 70
\reference{jay98} Jayawardhana, R., Fisher, S., Hartmann, L., Telesco, C., Pina, R. \& Giovanni, F.\ 1998, ApJ, 503, L79
\reference{jef61} Jeffreys H.\ 1961, Theory of probability, 3rd ed., Oxford University Press
\reference{jur93} Jura, M., Zuckerman, B., Becklin, E.E., \& Smith, R.C. 1993, \apj, 418, L37
\reference{jur95} Jura, M., Ghez, A.M., White, R. J., McCarthy, D.W., Smith, R.C., \& Martin, P.G., 1995, \apj, 445, 451
\reference{jur98} Jura, M., Malkan, M., White, R., Telesco, C., Pena \& Fisher, R.S. 1998, \apj, 505, 897
\reference{k98} Koerner, D.W., Ressler, M.E., Werner, M.W., \& Backman, D.E.,1998, \apjl, 503, L83
\reference{k00} Koerner, D.W., Jensen, E.L.N., Cruz, K.L., Guild, T.B. \& Gultekin, K.\ 2000, ApJ, 533, L37
\reference{k01} Koerner, D.W., Sargent, A.I. \& Ostroff, N.A.\ 2001, ApJ, 560, L181
\reference{kur93} Kurucz, R.\ 1993, ATLAS9 Stellar Atmosphere Programs 
and 2 km/s grid. Kurucz CD-ROM No. 13. Cambridge, Mass.: Smithsonian 
Astrophysical Observatory.
\reference{lp94} Lagage, P.O., \& Pantin, E., 1994, Nature, 369, 628
\reference{lch97} Lay, O.P., Carlstrom, J.E., \& Hills, R.E. 1997,\apj, 489, 917
\reference{ll03} Li, Aigen \& Lunine, J. I. \ 2003, \apj, 590, 368
\reference{lid04} Liddle, A.R.\ 2004, astro-ph/0401198
\reference{ld93}  Liou, J-C., Dermott, S. F., Abstracts for the IAU Symposium 160: Asteroids, Comets, Meteors 1993, 1993, p.191
\reference{liou96} Liou, J-C, Zook, H.A., Dermott, S.F.\ 1996, Icarus, 124, 429
\reference{luu02} Luu, J.X. \& Jewitt, D.C.\ 2002, Annual Review of Astronomy and Astrophysics, 40, 63
\reference{muk98} Mukherjee, S., Fiegelson, E.D., Babu, G.J., Murtagh, F., Fraley, C., Raftery, A.\ 1998, ApJ, 508, 314 
\reference{nak88} Nakano, T.\ 1988, MNRAS, 230, 551
\reference{nap00} Napier, W.M.\ 2000, MNRAS, 321, 463
\reference{ress94} Ressler, M.E., Werner, M.W.; Van Cleve, J., Chou, H.A.\ 1994, Experimental Astronomy, 3, 277
\reference{S99} Schneider, G., Smith, B.A., Becklin, E.E., Koerner, D.W., Meier, R., Hines, D.C., 
Lowrance, P.J., Terrile, R.J., Thompson, R.I., Rieke, M.\ 1999, ApJ, 513, 127
\reference{sit00}  Sitko, M.L., Lynch, D.K., Russell, R.W.\ 2000, \aj, 120, 2609
\reference{spies95} Spiesman, W.J., Hauser, M.G., Kelsall, T., Lisse, C.M., 
Moseley, S.H, Jr., Reach, W.T., Silverberg, R.F., Stemwedel, S.W., Weiland, J.L.\ 1995, \apj, 442, 662
\reference{stau95} Stauffer, J.R., Hartmann, L.W., Barrado y Navascues, D.\ 1995, \apj, 454, 910
\reference{T00} Telesco, C.M., Fisher, R.S., Piña, R.K., Knacke, R.F., Dermott, S.F., 
Wyatt, M.C., Grogan, K., Holmes, E.K., Ghez, A.M., Prato, L., 
Hartmann, L.W., \& Jayawardhana, R.\ 2000, \apj, 530, 329
\reference{torres95} Torres, C.A.O., Quast,G., de La Reza, R., Gregorio-Hetem,J., Lepine, J.R.D.\ 1995, \aj, 109, 2146
\reference{wahhaj03} Wahhaj, Z., Koerner, D. W., Ressler, M. E., Werner, M. W., Backman, D. E., Sargent, A. I. \ 2003, \apj, 584, 27
\reference{wein99} Weinberger, A.J., Becklin, E.E., Schneider, G., Smith, B.A., Lowrance, P.J., Silverstone, M.D., Zuckerman, B. \& Terrile, R.J.\ 1999, \apj, 525, L53
\reference{wein03} Weinberger, A.J., Becklin, E.E., Zuckerman, B.\ 2003, \apj, 584, 33
\reference{wil02} Wilner, D.J., Holman, M.J., Kuchner, M.J., Ho, P.T.P.\ 2002, 569, 115
\reference{W99} Wyatt, M.C., Dermott, S.F., Telesco, C.M., Fisher, R.S., Grogan, K., Holmes, E.K., \& Pi\~na, R.K.\ 1999, \apj, 527, 918
\reference{zb93} Zuckerman, B., \& Becklin, E.E., 1993, \apjl, 406, L25
\end{references}
\end{document}